\documentclass{aa}  

\usepackage{graphicx}
\usepackage{xcolor}
\usepackage{txfonts}
\usepackage{caption}
\usepackage{pdfpages}

\usepackage[colorlinks=true,
            linkcolor=blue,
            citecolor=blue,
            urlcolor=blue]{hyperref}

\begin{document}

   \title{Multi-spin stellar velocity maps of the most massive galaxies}
	\titlerunning{Multi-spin stellar velocity maps}
	\authorrunning{Krajnovi\'c et al. }

   \author{Davor Krajnovi\'c \inst{1},
          Eric Emsellem\inst{2,3}, 
         Roland Bacon\inst{3},
         Leindert A. Boogaard\inst{4},
         Peter M. Weilbacher\inst{1}, and
         Lutz Wisotzki\inst{1}
          }

   \institute{$^{1}$Leibniz-Institut f\"ur Astrophysik Potsdam (AIP), An der Sternwarte 16, D-14482 Potsdam, Germany (\email{dkrajnovic@aip.de})\\
             $^{2}$European Southern Observatory, Karl-Schwarzschild-Str. 2, 85748 Garching, Germany\\      
	    $^{3}$Univ Lyon, Univ Lyon1, ENS de Lyon, CNRS, Centre de Recherche Astrophysique de Lyon UMR5574, F-69230 Saint-Genis-Laval France\\
	    $^{4}$Leiden Observatory, Leiden University, PO Box 9513, NL-2300 RA Leiden, The Netherlands\\                   
             }

   \date{Received Dec 17, 2025; accepted Jan 30, 2026}

  \abstract
{We present stellar kinematics of the MUSE Most Massive Galaxies (M3G) Survey, comprising 25 galaxies brighter than $-25.7$ mag in the $K_s$-band and stellar mass above $\approx6\times10^{11}$ $M_\odot$. Galaxies are divided between the brightest cluster galaxies (BCGs) and lower-ranked (in brightness) galaxies (non-BCGs) in three rich galaxy clusters within the core of the Shapley super cluster. We find several velocity maps with rich kinematic structure, including multiple spin reversals within the region encompassing central two effective radii, typically associated with BCGs. Additionally, the majority of BCGs show rotation around the major-axis of the galaxy, at least in one of the visible velocity components. These kinematic structures are possible only if galaxies have non-axisymmetric shapes and contain several orbital families with both prograde and retrograde rotations. We argue that such velocity maps provide further evidence that the evolution of very massive galaxies is influenced by, possibly repeated, gas-poor major merging. There are six fast rotators in the M3G sample, all among non-BCGs, and typically ranked below the 3rd brightest galaxy in their clusters. Based on the properties of the $h_3$ Gauss-Hermite moment, fast rotation can be linked to the dominance of prograde rotating short-axis tubes in the orbital distribution. Slow rotators are BCGs or the second and sometimes third brightest galaxies, indicating that the galaxy mass (brightness) is not the only driver of low spin, but that the location within the local environment also plays a role. Slow rotators, as evidenced from their multi-spin velocity maps, require more complex orbital structures. Furthermore, some BCGs show kinematic evidence for a secondary component at larger radii, likely not in equilibrium with the main galaxy and possibly made of stars accreted from other cluster galaxies. Multi-spin velocity maps, low angular momentum and additional kinematic components highlight the difference in the evolutionary histories of BCGs (including 2nd ranked galaxies) and non-BCGs.  
}

   \keywords{galaxies: elliptical and lenticular, CD -- galaxies: formation, galaxies: evolution -- galaxies: kinematics and dynamics -- galaxies: structure
               }

   \maketitle

\section{Introduction}
\label{s:intro}

Galaxies with complex stellar kinematics have complex internal orbital structures. Predictions of stellar velocity maps \citep{1991AJ....102..882S, 1994MNRAS.271..924A} indicated that triaxial galaxies could sustain diverse velocity structures, because they can support multiple orbital families. In the separable potentials of the St\"ackel type there are four major orbital families: short-axis tubes, two types of long-axis tubes (inner and outer) and box orbits without net angular momentum \citep{1985MNRAS.216..273D}, visualised in \citet{1987ApJ...321..113S}\footnote{See also Chapter 13.4.1. in \citet{DynamicsAstrophysicsGalaxies} for a modern visualisation, available at \url{https://galaxiesbook.org/index.html}.}. In more realistic (non-rotating) triaxial potentials, irregular and resonant orbits, with descriptive names such as boxlets, bananas, saucers, fans and fish, are also possible \citep{1985MNRAS.216..467G, 1993ApJ...409..563S, 2014MNRAS.445.1065R}. Their spatial distributions are responsible for the observed shapes of galaxies \citep[e.g.][]{1979ApJ...232..236S}.

Non-axisymmetric galaxies can sustain velocity maps exhibiting kinematic twists, kinematically distinct cores (KDCs) and even counter-rotating cores. The origin of these kinematic structures was considered to be due to a combination of the projection effects and the existence of different orbital families \citep{1991AJ....102..882S, 1994MNRAS.271..924A}. However, except in a few cases where multiple slits were used to cover galaxies \citep[e.g.][]{1987ApJ...318..531G,1990dig..book..244W, 1999AJ....117..839S, 2002ApJ...578..787C} and where velocity maps were obtained through interpolation between the slits, long-slit observations were not able to determine and characterise kinematics structures on the velocity maps. 

The arrival of galaxy surveys with integral-field units \citep[IFUs;][]{2017_Bacon_book} at the beginning of the 21st century enabled a more comprehensive view of the kinematics of early-type galaxies (ETGs) \citep{2004MNRAS.352..721E}. The IFU maps of the mean velocity, V, and the velocity dispersion, $\sigma$, covering up to one effective radius ($R_e$) have several important advantages over long-slit kinematics. On the theory side, the maps can be directly connected to the velocity dispersion tensor and the tensor virial theorem \citep{2005MNRAS.363..937B, 2007MNRAS.379..418C}. Observationally, they enable the characterisation of kinematic properties. 

\citet{2011MNRAS.414.2923K} introduced five distinct classes based on a combination of visual inspection and harmonic analysis\footnote{See also figure 4 in \citet{2016ARA&A..54..597C} for a visual summary.}: {\it Group a}: no detectable rotation (3\%), {\it Group b}: irregular rotation (5\%), {\it Group c}: KDCs (7\%), {\it Group d}: counter-rotating disks (4\%) and {\it Group e}: disk-like rotation (80\%). The numbers in parenthesis refer to their frequencies in the magnitude limited ATLAS$^{\rm 3D}$\,sample \citep{2011MNRAS.413..813C} of local ETGs. Groups d and e are related as their galaxies are made of stellar disks and are nearly axisymmetric, even when supporting counter-rotation. Galaxies in Groups a--c do not have stellar disks \citep{2013MNRAS.432.1768K}. Furthermore, kinematic maps allow for the calculation of the projected angular momentum, as well as its proxy, the spin parameter $\lambda_{\rm R} \equiv \langle R|V| \rangle / R\sqrt{V^2 +\sigma^2}$ \citep{2007MNRAS.379..401E}, which can be calibrated to distinguish between galaxies with and without stellar disks \citep{2011MNRAS.414.2923K, 2011MNRAS.414..888E, 2016ARA&A..54..597C,2021MNRAS.505.3078V}.

Numerical simulations \citep{2014MNRAS.445.1065R, 2019MNRAS.489.2702F} and dynamical models constrained by IFU data \citep{2008MNRAS.385..647V, 2021MNRAS.508.4786D} are able to connect the observed complexity of stellar kinematics with the distribution of internal orbits. They show that velocity maps should be understood as luminosity-weighted superpositions of stellar motions stemming from different orbital families, including those with no net streaming. Orbital families can be co-spatial, with stars moving in the prograde and retrograde directions within any family, canceling or reinforcing streaming observed in the velocity maps. In some cases, velocity maps can appear as if there are multiple spin reversals \citep{2021MNRAS.508.4786D}. Crucially, the complexity of the observed kinematics can be related to the merger history \citep{2014MNRAS.444.3357N, 2017ApJ...850..144E, 2018MNRAS.473.1489L, 2021A&A...647A.103E, 2022MNRAS.509.4372L}.

In this work, we present the MUSE Most Massive Galaxies (M3G) Survey consisting of galaxies brighter than $-25.5$ in the $K$-band. We focus on the survey design and observations, presenting the kinematics and an analysis of the velocity maps. Such galaxies are rare in the nearby Universe and their stellar kinematics is not yet well mapped, in spite of recent works \citep{2008MNRAS.391.1009L, 2014ApJ...795..158M, 2022MNRAS.515.1104L}. An expectation at the time of the M3G survey definition was that they would have little or no rotation, belonging to Groups a - c. If there were any exceptional kinematic structures, they would be confined to their cores, within a few kpc at most. The primary purpose of this paper is to demonstrate and quantify the surprising richness of observed features in M3G velocity maps. As a result, it confirms their origin through the superposition of stellar motions belonging to different orbital families. 

This paper is organised as follows: in Section~\ref{s:obs} we present the M3G sample, MUSE observations and data reduction. In Section~\ref{s:kin} we describe the extraction of stellar kinematics, while in Section~\ref{s:kina} we present the kinematic analysis and the main results of the paper. Discussion and conclusions are presented in Sections~\ref{s:disc} and~\ref{s:con}, respectively. The kinematic maps of M3G galaxies and other supporting material are attached in the Appendices. We use the standard cosmology and H$_0=70$ km/s/Mpc.


   \begin{figure*}
    \includegraphics[width=0.95\columnwidth]{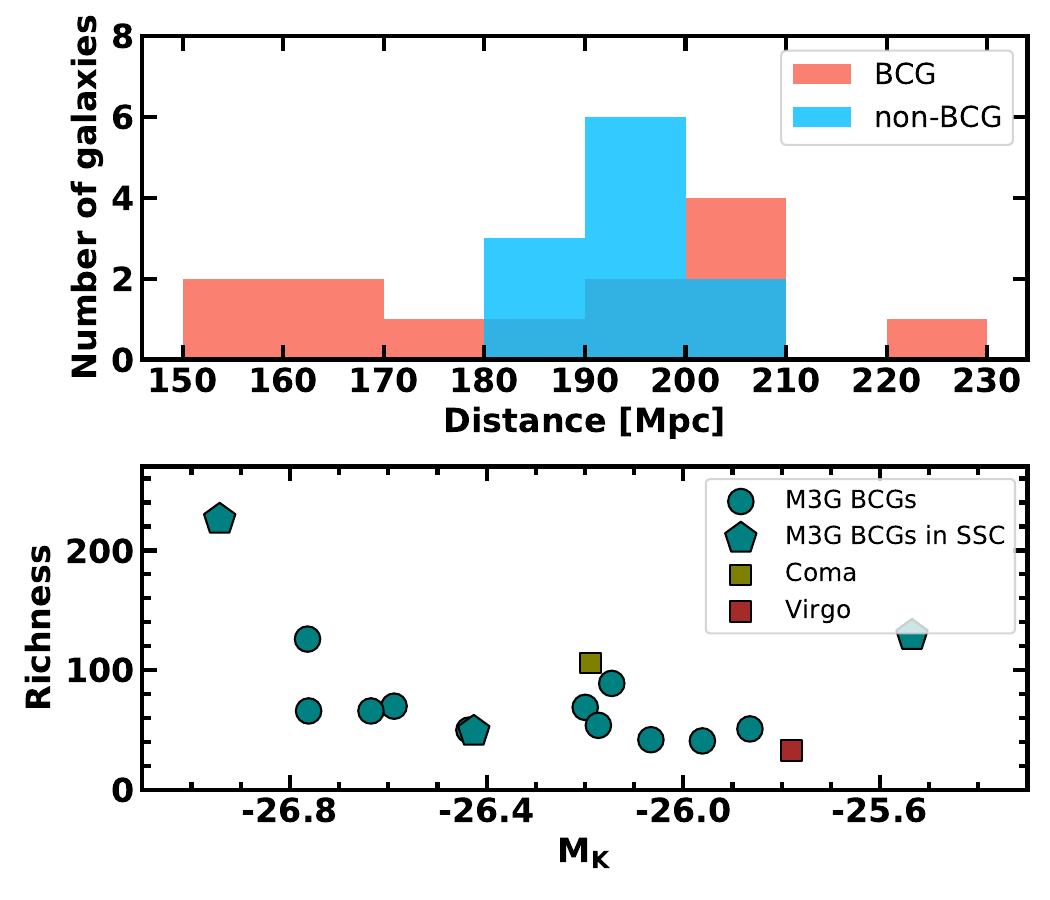} 
    \includegraphics[width=0.95\columnwidth]{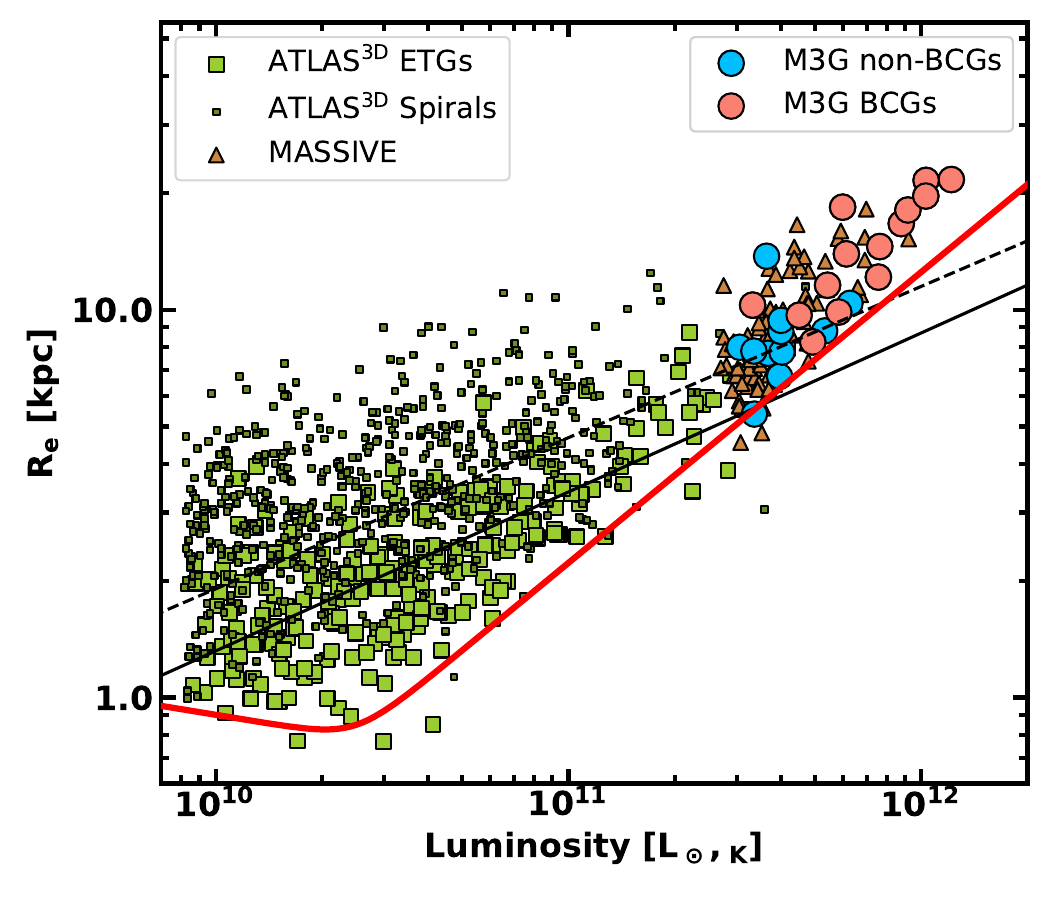}
      \caption{{\it Left top:}  A histogram of the distances to the BCGs (orange) and the satellites galaxies (blue). {\it Left bottom:} Richness of the host galaxy cluster vs the absolute $K_{s}$ magnitude of the corresponding BCG. BCGs in the Shapley Super Cluster are shown as pentagrams (Abell\,3558, Abell\,3556 and Abell\,3562 in order of decreasing brightness of the BCGs). Virgo and Coma cluster (squares) are added as references for the ATLAS$^{\rm 3D}$ and the MASSIVE Surveys. {\it Right:} The size -- luminosity relation for the M3G sample galaxies, split in BCGs (red) and satellites (blue). Galaxies from two other surveys are added for comparison: ETGs from the ATLAS$^{\rm 3D}$ Survey (light green large squares), spirals from the ATLAS$^{\rm 3D}$ Survey parent sample (dark green small squares), and ETGs from the MASSIVE survey (brown triangles). ATLAS$^{\rm 3D}$ sample is a magnitude-limited sample within 42 Mpc, while the galaxies in the MASSIVE are the brightest galaxies within 100 Mpc. The solid and dashed black lines are relations for fast rotators and S0-Sa galaxies from the ATLAS$^{\rm 3D}$ survey, respectively. The red line denotes the so-called `zone of avoidance'. All three relations are taken from \citet{2011MNRAS.413..813C}. }
         \label{f:sample}
   \end{figure*}

\section{The galaxy sample, MUSE observations and data reduction}
\label{s:obs}

The M3G sample, MUSE observations and the data reduction were already partially presented in \citet{2018MNRAS.477.5327K}. Here we repeat the pertinent details and point to the new developments since the publication of that paper. Some of these developments were already discussed in \citet{2021A&A...649A..63P} and \citet{2021MNRAS.508.4786D,2024MNRAS.530.3278D} and are summarised here. 

\subsection{The properties of M3G sample galaxies}
\label{ss:sample}

The idea behind the M3G sample was to explore the spectroscopic properties of galaxies in a way not possible by previous surveys: to observe very massive galaxies that are rare in the nearby universe beyond their effective radius. The sample was in practice selected by imposing the following criteria: 

\begin{itemize}
\item galaxy sizes were required to match the MUSE FoV such that  $2R_e$ are covered within a single pointing, 
\item galaxies were required to be brighter than $-25.5$ magnitude, corresponding to $\sim5\times10^{11}$ $M_\odot$ \citep[following Eq. 2 in][]{2013ApJ...778L...2C}, the upper limit brightness and mass ranges for the ATLAS$^{\rm 3D}$ sample \citep{2011MNRAS.413..813C},
\item galaxies were supposed to be in dense environments, and in particular rich clusters (richer than the Virgo cluster, also covered by the ATLAS$^{\rm 3D}$ sample),
\item most galaxies of this brightness limit were expected to be the brightest cluster galaxies (BCGs), but a significant fraction of galaxies was required to belong to non-BCGs, and
\item the spatial resolution was required to be better than 1 kpc, as this is the typical size of KDCs in massive early-type galaxies \citep{2011MNRAS.414.2923K}. Assuming natural seeing conditions at the VLT, this constraint limited the distance to about 250 Mpc for the selection of massive galaxies.
\end{itemize}

We searched for candidate galaxies in the core of the Shapley Super Cluster \citep[SSC;][]{1930BHarO.874....9S,2010MNRAS.402..753M, 2015MNRAS.446..803M, 2018MNRAS.481.1055H}, focusing on Abell\,3556, 3558 and 3562, the three largest clusters within the overall supercluster structure. The magnitude selection provided 14 galaxies, among which are also BCGs of the three clusters. We further matched this SSC sub-sample with 11 BCGs based on the sample of galaxies presented in \citet{2003AJ....126.2717L}, which itself is based on the survey of \citet{1989ApJS...70....1A} clusters by \citet{1994ApJ...425..418L} and \citet{1995ApJ...440...28P}. The galaxies were located in the southern hemisphere (declination $< 5\degr$), in clusters that have the \citet{1958ApJS....3..211A} richness\footnote{Richness is defined as the number of galaxies up to two magnitudes fainter than the third brightest galaxy, within the Abell radius.} larger than the Virgo Cluster (33), that match the magnitude and distance cut of the SSC subsample. As emphasised by \citet{1989ApJS...70....1A}, richness is not a substitute for local luminosity functions, and we use it only as a relative measure of environments denser than those covered by the ATLAS$^{\rm 3D}$ Survey \citep{2011MNRAS.416.1680C}. The three BCGs in the SSC sample are also part of the \citet{2003AJ....126.2717L} sample, and therefore all 14 BCGs have HST imaging, while only two non-BCGs were imaged with HST. 

The final M3G sample, therefore, is made of 25 galaxies, 14 BCGs and 11 members of the three main galaxy clusters in the core of the SSC. We will refer to these galaxies as {\it non-BCGs}, and they comprise at least the second and the third brightest galaxies in their respective clusters, as well as galaxies ranked to lower brightnesses. Based on galaxies in nearby clusters (e.g. Virgo or Coma cluster), the second brightest galaxies are often very similar to the nominal BCGs, and for some interpretations this needs to be kept in mind \citep{2014ApJ...797...82L}. The general properties of the sample galaxies are presented in Table~\ref{t:m3g}. The emission-line properties and the dust content were discussed in \citet{2021A&A...649A..63P}, while the stellar population and elemental abundances were presented in \citet{2024MNRAS.530.3278D}. We do not repeat them here. 

Three galaxies in the M3G sample have incorrect designation in previous papers \citep{2018MNRAS.477.5327K, 2021A&A...649A..63P, 2024MNRAS.530.3278D}. They are PGC\,097958, PGC\,099188 and PGC\,099522. The correct designations for PGC\,097958 and PGC\,099522 are LEDA\,097958 (in Abell\,3558) and LEDA\,099522 (in Abell\,3562), because the Catalogue of Principle Galaxies has 73197 official entries \citep{1989A&AS...80..299P}, while catalogue entries with larger number are designated as LEDA. They are listed in the Lyon-Medon Extragalactic Database \citep{1988ESOC...28..435P}, and services such as SIMBAD\footnote{\url{https://simbad.cds.unistra.fr/simbad/}} easily resolve between PGC and LEDA catalog names. The galaxy previously named PGC\,099188 is actually PGC\,047548 (in Abell\,3562), with coordinates RA=13 31 27.52 DEC=-31 49 14.60 (J2000), also designated as ShaSS403013239 \citep{2015MNRAS.453.3685M}. PGC\,099188 (LEDA\,099188), or ShaSS403010664 is a faint galaxy at z=0.21 \citep{2018MNRAS.481.1055H}. In order to highlight this inaccuracy, but to simplify the cross-correlation with previous M3G sample papers and designation of the MUSE data set in the ESO archive\footnote{Using FITS keyword OBJECT for searching, and via coordinates by resolving the name.} we will nevertheless keep the designation of PGC099188* (highlighted by a star) for PGC\,047548.

We made use of the 2MASS $K_s$-band magnitudes (\texttt{k\_m\_ext}, the extrapolated K$_{s}$  band 2MASS magnitude) and size measurements from the All Sky Extended Source Catalogue \citep[XSC;][]{2000AJ....119.2498J, 2006AJ....131.1163S}. The $R_e$ of our galaxies was estimated as the mean of the 2MASS measured effective radii in $J$, $H$ and $K_{s}$ bands, $R_e$ = mean(\texttt{j\_r\_eff}, \texttt{h\_r\_eff}, \texttt{k\_r\_eff}). We compared these $R_e$ to $R_e^{maj}$ obtained as the major axis of the half-light isophote \citep{2013MNRAS.432.1709C}, by using the uniform r-band VST images of galaxies in the SSC \citep{2015MNRAS.446..803M} to construct Multi-Gaussian Expansion (MGE) models\footnote{We used the software based on \citet{2002MNRAS.333..400C} that is available from \url{https://pypi.org/project/mgefit/}} \citep{1994A&A...285..723E}, and estimate R$_e^{maj}$.  Comparing the calculated $R_e^{maj}$ with the $R_e$ for the 14 SSC galaxies, we obtained a linear relation, with the slope $b=1.0\pm 0.1$. As there is no uniform VST imaging for the full sample to facilitate a uniform calculation of $R_e^{maj}$, we adopted the 2MASS-based half-light radii, as defined above. 

A major source of uncertainty for the sample properties is the distance measurements. Only about 1/3 of the sample has known distance estimates, and those are based on scaling relations, such as the fundamental plane. Therefore, we estimated the distance as follows. Firstly, we collected all available distances from the NASA Extragalactic data base\footnote{Available at: \url{https://ned.ipac.caltech.edu}}. When there were two or more distance estimates, which were not strongly discrepant, we used the average value. For galaxies that did not have distance estimates, we collected the redshift measurements and used the Hubble law to obtain the distance given the redshift:
\begin{equation}
D \simeq \frac{c}{H_0}\frac{c (z+1)^2 -1}{c (z+1)^2 +1},
\end{equation}
Finally, we required that all galaxies within the same cluster (SSC galaxies) have the same distance, ignoring the distribution of galaxies within a cluster. We calculated the SSC clusters distances as the median of the available distances of all cluster galaxies present in our sample. Based on the mean recession velocities \citep{2018MNRAS.481.1055H}, the distances to Abell\,3556, Abell\,3558 and Abell\,3562 are 200 Mpc, 202 Mpc and 205 Mpc, which is up to 10\%\, different from our final estimates (Table~\ref{t:m3g}). \citet{2018MNRAS.481.1055H} report a large spread of velocities for the three SSC clusters ($\sigma_c=$ 628km/s, 1007km/s and 769 km/s, respectively), and a distance range of about 20-30 Mpc within the cluster, also amounting to about 10\% with respect to the cluster mean distance. Overall, we estimate the accuracy of the distances to be about 10\%, but this has no influence on the results of this paper. The distribution of distances is shown in Fig.~\ref{f:sample}: SSC galaxies are located between 180 and 200 Mpc, while the BCGs have a somewhat larger spread. The adopted distances mean that 1\arcsec\,is between 0.73 and 1.12 kpc for the M3G sample. The new distances and the effective radii take precedence over those published in previous M3G publications. 

Fig.\ref{f:sample} shows where the M3G targets stand with respect to local samples of early-type galaxies investigated with IFU spectroscopy. M3G galaxies are found on the apex of the luminosity-size relation, reaching luminosities of $10^{12} L_\odot$ and half-light sizes larger than 20 kpc. They are thus similar to the galaxies in the MASSIVE sample \citep{2014ApJ...795..158M}, with a few BCGs being somewhat more luminous. The non-BCGs in the sample have, as expected, lower luminosities and smaller sizes, but can still be considered extreme systems. 

In terms of the environment, M3G galaxies are found in clusters with richnesses between those of the Virgo and Coma clusters. Abell\,3558 is by far the richest cluster, containing almost a factor of two more galaxies than other probed clusters, and contains the brightest galaxy in the sample (PGC\,047202). Abell\,3562, while being the second richest cluster in the sample, harbours the faintest BCGs (PGC\,047752). Two other galaxies from this cluster, PGC\,099188* and PGC\,099522, were found to actually have larger 2MASS apparent luminosities than PGC\,047752 (10.7 mag and 10.6 mag, compared to 10.8 mag, respectively). This could be driven by the presence of nearby overlapping satellites: they have significantly different systemic velocities, but fall within the MUSE FoV and strongly influence the light distributions within the central 10\arcsec. Additionally, Abell\,3562 experienced a recent supersonic encounter with the SC 1329-313 group \citep{2004ApJ...611..811F}. This is relevant as the centre of the X-ray emission in the vicinity of SC 1329-313 group coincides with the location of PGC\,47548, while the similarly bright PGC\,047590 is located in the vicinity along the same X-ray filament emission. A brighter X-ray emission source is centred on PGC\,047752 identified as the BCG of Abell\,3562 by \citet{1995ApJ...440...28P}. \citet{2018MNRAS.481.1055H} assigns both PGC\,47548 and PGC\,047590 to Abell\,3562 based on velocity-related arguments. Therefore, M3G members of Abell\,3562 have similar apparent 2MASS $K_s$-band magnitudes (10.7, 10.6, 10.8 and 10.6 mag in the order of the increasing PGC number), where the faintest can be taken as the local BCG, while the brightness of others might be overestimated by the presence of companions.

We note that the differences in the derived absolute magnitudes for M3G Abell\,3562 galaxies are less than 0.3 mag, a level sometimes used to consider the second-ranked galaxy a close rival to the local BCG \citep{2014ApJ...797...82L}. In this respect, all M3G galaxies of Abell\,3562 could be seen as close rivals for the BCG place. We will nevertheless keep PGC\,047752 (also known as ESO 444-G72) as the BCG of Abell\,3562. PGC\,099188*, PGC\,047752 and PGC\,099522 are assigned the rank of the second, third and fourth galaxies in the cluster, and we remind the reader to keep in mind the connection with SC 1329-313 group for PGC\,099188* and PGC\,047590, and a possibly overestimated luminosity for PGC\,099522.


\begin{table*}
\small
\caption{General properties of M3G galaxies}             
\label{t:m3g}   
\setlength{\tabcolsep}{5pt}   
\begin{tabular}{llccccccccccccc}
   \hline
    \noalign{\smallskip}
\multicolumn{1}{c}{galaxy} &
\multicolumn{1}{c}{Abell} &
\multicolumn{1}{c}{BCG} &
\multicolumn{1}{c}{D} &
\multicolumn{1}{c}{$M_K$} &
\multicolumn{1}{c}{Rich} &
\multicolumn{1}{c}{$R_{\rm e}$} &
\multicolumn{1}{c}{$R_{\rm e}$} &
\multicolumn{1}{c}{$\sigma_{\rm e}$} &
\multicolumn{1}{c}{$\lambda_{\rm Re}$} &
\multicolumn{1}{c}{$\epsilon_{\rm Re}$} &
\multicolumn{1}{c}{$\Delta k_0$} &
\multicolumn{1}{c}{kin type} &
\multicolumn{1}{c}{multi spin} &
\multicolumn{1}{c}{Group} \\
\multicolumn{1}{c}{} &
\multicolumn{1}{c}{} &
\multicolumn{1}{c}{} &
\multicolumn{1}{c}{Mpc} &
\multicolumn{1}{c}{} &
\multicolumn{1}{c}{} &
\multicolumn{1}{c}{\arcsec} &
\multicolumn{1}{c}{kpc} &
\multicolumn{1}{c}{km s$^{-1}$} &
\multicolumn{1}{c}{} &
\multicolumn{1}{c}{} &
\multicolumn{1}{c}{km s$^{-1}$} &
\multicolumn{1}{c}{} &
\multicolumn{1}{c}{} &
\multicolumn{1}{c}{} \\
\multicolumn{1}{c}{(1)} &
\multicolumn{1}{c}{(2)} &
\multicolumn{1}{c}{(3)} &
\multicolumn{1}{c}{(4)} &
\multicolumn{1}{c}{(5)} &
\multicolumn{1}{c}{(6)} &
\multicolumn{1}{c}{(7)} &
\multicolumn{1}{c}{(8)} &
\multicolumn{1}{c}{(9)} &
\multicolumn{1}{c}{(10)} &
\multicolumn{1}{c}{(11)} &
\multicolumn{1}{c}{(12)} &
\multicolumn{1}{c}{(13)} &
\multicolumn{1}{c}{(14)} &
\multicolumn{1}{c}{(15)} \\    \noalign{\smallskip} 
    \hline \hline
    \noalign{\smallskip}
PGC003342   & 0119 & 1   & 165 & -26.20  & 69  & 17.41 & 13.9    & $262 \pm 2$  & 0.037        & 0.252      & 8   & NRR      & MS-2L        & b$^\dagger$        \\
PGC004500   & 0168 & 1   & 203 & -26.15 & 89   & 10.02 & 9.9     & $252 \pm 2$  & 0.038        & 0.110      &12   & NRR      & MS-1         & a         \\
PGC007748   & 0295 & 1   & 178 & -25.86 & 51   & 11.19 & 9.7     & $262 \pm 2$  & 0.040        & 0.283      &6    & NRR      & MS-2L        & c        \\
PGC015524   & 0496 & 1   & 151 & -26.44 & 50   & 19.87 & 14.5    & $274 \pm 2$  & 0.041        & 0.194      &51   & NRR      & MS-1L        & b        \\
PGC018236   & 3376 & 1   & 195 & -26.07 & 42   & 12.23 & 11.6    & $273 \pm 2$  & 0.070        & 0.314      &17   & NRR      & MS-1         & b        \\
PGC019085   & 3395 & 1   & 202 & -26.17 & 54   & 18.68 & 18.4    & $257 \pm 2$  & 0.038        & 0.520      &32   & NRR      & MS-3L        & c          \\
PGC043900   & 3528 & 1   & 225 & -26.59 & 70   & 15.25 & 16.7    & $369 \pm 2$  & 0.032        & 0.241      &20   & NRR      & MS-4L        & a$^\dagger$         \\
PGC046785   & 3556 & 2   & 200 & -26.23 & 49   & 10.69 & 10.4    & $321 \pm 2$  & 0.108        & 0.066      &7    & RR       & MS-2         & e      \\
PGC046832   & 3556 & 1   & 200 & -26.43 & 49   & 12.49 & 12.1    & $308 \pm 2$  & 0.051        & 0.323      &17   & NRR      & MS-5L        & c        \\
PGC046860   & 3556 & 3   & 200 & -25.73 & 49   & 6.93  & 6.7     & $278 \pm 2$  & 0.332       & 0.333      &16   & RR       & MS-1        & e       \\
PGC047154   & 3558 & 2   & 193 & -26.05 & 226  & 9.42  & 8.8     & $320 \pm 2$  & 0.043        & 0.160      &12   & NRR      & MS-1         & b       \\
PGC047177   & 3558 & 4   & 193 & -25.64 & 226  & 8.25  & 7.7     & $284 \pm 2$  & 0.360        & 0.368      &25   & RR       & MS-1         & e        \\
PGC047197   & 3558 & 3   & 193 & -25.74 & 226  & 8.31  & 7.8     & $301 \pm 2$  & 0.035        & 0.231      &11   & NRR      & MS-2L        & c       \\
PGC047202   & 3558 & 1   & 193 & -26.94 & 226  & 23.14 & 21.7    & $311 \pm 2$  & 0.114        & 0.264      &124  & NRR      & MS-1         & b        \\
PGC047273   & 3558 & 7   & 193 & -25.44 & 226  & 8.55  & 8.0     & $254 \pm 2$  & 0.304        & 0.239      &26   & RR       & MS-1         & e      \\
PGC047355   & 3558 & 6   & 193 & -25.54 & 226  & 8.36  & 7.8     & $253 \pm 2$  & 0.656        & 0.238      &7    & RR       & MS-1         & e      \\
PGC047590   & 3562 & 2   & 183 & -25.73 & 129  & 9.88  & 8.8     & $279 \pm 2$  & 0.034        & 0.212      &10   & NRR      & MS-2L        & c      \\
PGC047752   & 3562 & 1   & 183 & -25.53 & 129  & 11.55 & 10.3    & $249 \pm 2$  & 0.061        & 0.215      &33   & NRR      & MS-4L       & c        \\
PGC048896   & 3571 & 1   & 162 & -26.76 & 126  & 27.42 & 21.5    & $329 \pm 2$  & 0.041        & 0.503      &7    & NRR      & MS-2L       & c        \\
PGC049940   & 1836 & 1   & 154 & -25.96 & 41   & 11.04 & 8.3     & $286 \pm 2$  & 0.063        & 0.135      &17   & NRR      & MS-1         & b        \\
PGC065588   & 3716 & 1   & 231 & -26.64 & 66   & 16.08 & 18.1    & $264 \pm 2$  & 0.068        & 0.089      &15   & NRR      & MS-2L        & b$^\dagger$       \\
PGC073000   & 4059 & 1   & 204 & -26.76 & 66   & 19.76 & 19.6    & $278 \pm 2$  & 0.046        & 0.298      &35   & NRR      & MS-1L        & b        \\
PGC097958  & 3558 & 5   & 193 & -25.55 & 226  & 5.75  & 5.4     & $282 \pm 2$  & 0.350        & 0.289      &10   & RR       & MS-1         & e      \\
PGC099188*   & 3562 & 3   & 183 & -25.63 & 129  & 15.44 & 13.7    & $231 \pm 4$  & 0.060        & 0.128      &24   & NRR      & MS-2L        & c        \\
PGC099522   & 3562 & 4   & 183 & -25.74 & 129  & 10.55 & 9.4     & $225 \pm 2$  & 0.068        & 0.247      &35   & NRR      & MS-1         & b        \\
 \hline                  
\end{tabular}
\\
Notes: Column (1): name of the galaxy. Note that PGC099188* is actually PGC\,047548 (see Section~\ref{ss:sample}). Column (2): number of the Abell cluster. Column (3): 1 -- galaxy is a BCG, 2 or higher number -- the luminosity rank of the galaxy within its cluster, from the 2nd brightest galaxy to fainter galaxies. Galaxies with number larger than 1 are members of one of the clusters in the SSC (also denoted as non-BCG in the text). Column (4): distance to the galaxy, as described in Section ~\ref{ss:sample}. Column (5): absolute magnitude in $K_s$-band based on the \texttt{k\_m\_ext} keyword of 2MASS XSC catalog and the distance from Column (4). Column (6): Richness of the cluster. Column (7): effective radius in arc seconds, estimated as the mean of \texttt{j\_r\_eff, h\_r\_eff} and \texttt{k\_r\_eff} values from the 2MASS XSC catalog. Column (8): effective radius in kpc. Column (9): effective velocity dispersion extracted within the elliptical aperture with semi-major axis equal to $R_{\rm e}$. Column (10): $\lambda_{Re}$ (uncertainties are less then 0.005). Column (11): ellipticity of the galaxy within $R_{\rm e}$. Column (12): Variation (max-min) of the $k_0$ coefficient across the FoV.  Column (13): kinematic type based on \textsc{kinemetry} analysis, where NRR = non-regular rotator and RR = regular rotator.  Column (14): number of the spin reversals visible on the velocity map. ``L" indicates the existence of a component rotating around the long-axis. Column (15): kinematic group (within $1R_e$): see Sections~\ref{s:intro} and \ref{sss:nonR}, or \citep{2011MNRAS.414.2923K}. Entries with $^\dagger$ would be sorted in Group c, if 2$R_e$ region was used for the classification. 

\end{table*}

\subsection{MUSE observations}
\label{ss:obs}

The M3G sample was observed as part of the MUSE Guaranteed Time Observations (GTO) during ESO periods 94-99 spanning three years (2014-2017)\footnote{The MUSE observations were taken within the following observing programmes: 094.B-0592, 095.B-0127, 096.B-0062, 097.B-0776, 098.B-0240, 099.B-0148, 099.B-0242 and 0102.A.0327.}. Each galaxy was observed using MUSE in Wide Field Mode without Adaptive Optics and a nominal wavelength range (480 -- 930~nm). Observing Blocks (OB) were split into four on-target and two sky observations within an OSOOSO sequence (O is Object, S is Sky). Consecutive on-target observations were rotated by 90$^{\rm o}$ and dithered to reduce systematics between the individual spectrographs. The sky exposures (120s) were pointing several arc minutes away from the galaxy to areas free from cluster galaxies (and foreground stars). No other special calibration exposures were used, except routine standard stars observed once per night. 

\begin{figure}
    \includegraphics[width=\columnwidth]{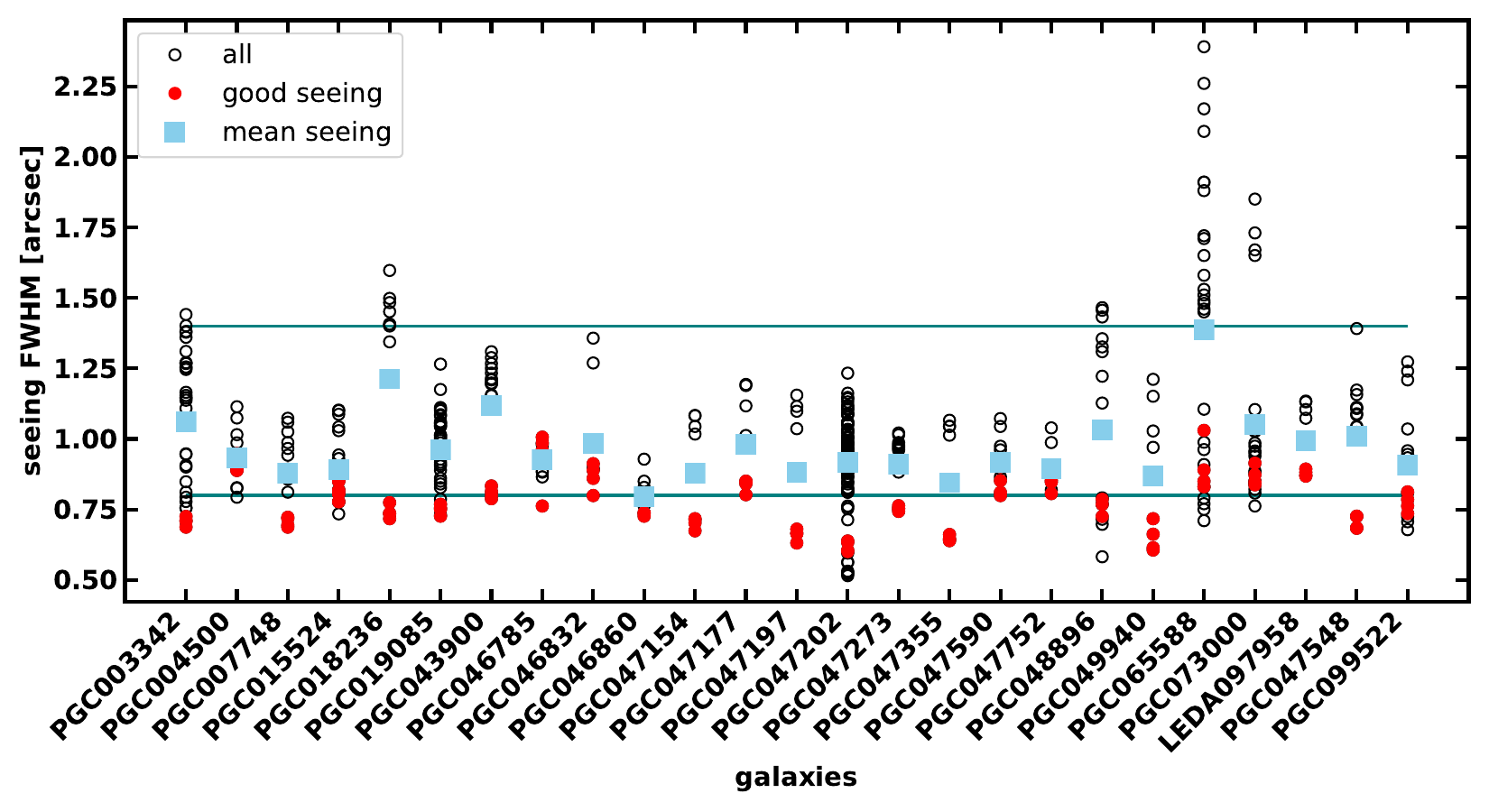} 
      \caption{Distribution of the VLT auto-guider seeing measurements, averaged over each exposure taken with MUSE for all M3G galaxies. Horizontal lines denote the two seeing limits imposed on the programme. Blue squares show the mean value of all observations for each galaxy. Red circles are exposures which were targeted to be done in ``good seeing'' conditions (FWHM<0.8$^{\prime\prime}$). }
         \label{f:seeing}
   \end{figure}

The M3G survey was assigned to be a "poor seeing" programme within the GTO projects, meaning that it was executed when the FWHM of the ambient seeing would exceed 1\arcsec, but be no worse than $1\farcs4$. A fraction of GTO time was further allocated for relatively short on-target exposures ($\approx 1800$s) using "good seeing" (GS: FWHM<$0\farcs8$) conditions to resolve the central core structures (smaller than 1~kpc). In practice, exposures within one OB exhibit strongly varying seeing conditions. In Fig.~\ref{f:seeing}, we show the distribution of recorded seeing values by the VLT auto-guider. The resulting mean seeing of the survey is $0.93\arcsec$, with 28\% of the sample galaxies having the mean seeing poorer than 1\arcsec. This is better than expected given the observational strategy. In several cases, however, the GS exposures did not achieve the target specification, with 44\% galaxies having seeing worse than the targeted $0\farcs8$, but the median seeing of the GS data cubes is $0\farcs75$. The seeing values for each galaxy are listed in Table~\ref{t:obs}. In this work, we combine all available exposures to increase the signal and only occasionally use the GS cubes to verify our kinematic analysis in the nuclei (Section~\ref{sss:nonR}), but do not present them here. 

Total exposure times for each galaxy ranged from 2 to 6 h. Table~\ref{t:obs} also lists the limiting surface brightness magnitudes reached at the edge of the extracted kinematic maps (see Section~\ref{s:kin}). The magnitudes were extracted from the MUSE reconstructed images using the SDSS $r$-band filter. The average limiting surface brightness reached is 23.21 mag/\arcsec$^2$(AB magnitudes) with a standard deviation of 0.38 mag/\arcsec$^2$, showcasing the achieved uniformity of observations.

Each galaxy was observed by placing its nucleus at the centre of the MUSE FoV. For the central galaxy of Abell\,3558 and the brightest in the sample, PGC\,047202, we built a $2\times2$ mosaic from the nominal "poor-seeing" exposures, each time placing its nucleus in one MUSE corner, in addition to "good-seeing" exposures with the nucleus placed in the centre on the MUSE FoV. 

The quality of the individual exposures is typically high. A few individual exposures were not completed or, upon inspection, found to be of insufficient quality to be included for the final scientific analysis. In some cases, exposures had a reduced flux (i.e. down by 20-40\% for PGC\,047202 in the SW quadrant, see Fig.~\ref{f:kinem}), due to passing clouds, and in two instances for PGC\,065588, satellite trails were also recorded. These observations were nevertheless kept and combined with the rest, resulting in still valuable albeit not ideal data. Exposures were rejected from the final creation of the data cubes if they had been aborted, usually due to the loss of a guide star (PGC\,03342, PGC\,15524, PGC\,047202). One OB belonging to PGC\,048896 was observed through clouds and resulted in no exploitable data, while two PGC\,019085 OBs suffered from incorrect offsets when moving between on-target and sky locations.


\begin{table}
\small
\caption{Observations of M3G galaxies}             
\label{t:obs}      
\setlength{\tabcolsep}{4pt}   
\begin{tabular}{lccccccc}
   \hline
    \noalign{\smallskip}
\multicolumn{1}{c}{name} &
\multicolumn{1}{c}{T} &
\multicolumn{1}{c}{$N_{\rm exp}$} &
\multicolumn{1}{c}{$N_{\rm com}$} &
\multicolumn{1}{c}{$T_{\rm GS}$} &
\multicolumn{1}{c}{PSF} &
\multicolumn{1}{c}{PSF$_{\rm GS}$} &
\multicolumn{1}{c}{m$_r$} \\
\multicolumn{1}{c}{} &
\multicolumn{1}{c}{[s]} &
\multicolumn{1}{c}{} &
\multicolumn{1}{c}{} &
\multicolumn{1}{c}{[s]} &
\multicolumn{1}{c}{$^{\prime\prime}$} &
\multicolumn{1}{c}{$^{\prime\prime}$} &
\multicolumn{1}{c}{AB mag} \\
\multicolumn{1}{c}{(1)} &
\multicolumn{1}{c}{(2)} &
\multicolumn{1}{c}{(3)} &
\multicolumn{1}{c}{(4)} &
\multicolumn{1}{c}{(5)} &
\multicolumn{1}{c}{(6)} &
\multicolumn{1}{c}{(7)} &
\multicolumn{1}{c}{(8)} \\    
\noalign{\smallskip} 
    \hline \hline
    \noalign{\smallskip}
PGC003342 & 16200 & 33 & 32 & 1080 & 1.06 & 0.71 & 23.15\\
PGC004500 & 5040 & 12 & 12 & 720 & 0.93 & 0.92 & 23.26\\
PGC007748 & 6390 & 13 & 13 & 720 & 0.88 & 0.71 & 23.91\\
PGC015524 & 9000 & 20 & 16 & 720 & 0.89 & 0.81 & 23.04\\
PGC018236 & 4680 & 12 & 12 & 1080 & 1.21 & 0.74 & 23.54\\
PGC019085 & 17100 & 39 & 29 & 1440 & 0.96 & 0.75 & 23.23\\
PGC043900 & 6840 & 16 & 16 & 720 & 1.12 & 0.81 & 23.44\\
PGC046785 & 8280 & 16 & 16 & 720 & 0.93 & 0.93 & 22.93\\
PGC046832 & 3240 & 8 & 8 & 720 & 0.98 & 0.86 & 23.14\\
PGC046860 & 2880 & 8 & 8 & 720 & 0.80 & 0.75 & 23.13\\
PGC047154 & 3240 & 8 & 8 & 720 & 0.88 & 0.70 & 23.67\\
PGC047177 & 2520 & 8 & 8 & 720 & 0.98 & 0.84 & 22.94\\
PGC047197 & 2880 & 8 & 8 & 720 & 0.88 & 0.66 & 23.10\\
PGC047202 & 59670 & 103 & 101 & 1440 & 0.92 & 0.62 & 22.74\\
PGC047273 & 7200 & 16 & 16 & 720 & 0.91 & 0.75 & 23.39\\
PGC047355 & 3240 & 8 & 8 & 1080 & 0.84 & 0.65 & 23.08\\
PGC047590 & 5940 & 12 & 12 & 1980 & 0.92 & 0.84 & 23.43\\
PGC047752 & 3960 & 8 & 8 & 1440 & 0.89 & 0.85 & 22.89\\
PGC048896 & 12480 & 24 & 20 & 1440 & 1.03 & 0.73 & 22.05\\
PGC049940 & 2160 & 8 & 8 & 720 & 0.87 & 0.65 & 22.70\\
PGC065588 & 16830 & 30 & 29 & 1080 & 1.39 & 0.90 & 23.29\\
PGC073000 & 15120 & 24 & 24 & 2520 & 1.05 & 0.87 & 23.60\\
PGC097958 & 2880 & 8 & 8 & 1440 & 0.99 & 0.88 & 23.52\\
PGC099188* & 8640 & 16 & 16 & 1440 & 1.01 & 0.70 & 23.83\\
PGC099522 & 6480 & 16 & 16 & 720 & 0.91 & 0.77 & 23.14\\
\hline                  
\end{tabular}
\\
Notes: Column (1): galaxy name. Column (2): total on-source exposure time of the final data cubes. Column (3): number of individual exposures. Column (4): Number of individual exposures used for the creation of the final data cube. Column (5): total exposure time of ``good seeing" exposures. Column (6): mean seeing (FWHM) of all exposures as reported by the telescope auto-guider. Column (7): mean seeing  (FWHM) of ``good seeing" exposures (see text for details). Column (8): limiting surface brightness (SDSS $r$-band AB magnitudes) reached at the edge of the kinematic maps shown in Fig.~\ref{f:kinem}. 

\end{table}

\subsection{Data reduction}
\label{ss:drs}

M3G galaxies were observed during 15 observing runs spread over six semesters, with data reduction proceeding on individual exposures as they were completed. This means that we used different versions (v1.2 to v1.7) of the MUSE data reduction pipeline \citet{2020A&A...641A..28W} as they became available. We verified that this has not significantly impacted the reduction products and the results presented in this paper. All reductions followed the standard MUSE pipeline steps, from master calibration files (bias, flat fields, trace tables), including the wavelength calibration and the line-spread function (LSF) estimates for each spectral layer. In a number of cases, we used twilight flats (if observed on the same night). Instrument geometry and astrometry were produced at each observing run by the GTO team. We paid particular attention to using the illumination flats as close to the individual on-target exposures as possible. 

Removing the sky contribution is non-trivial when objects fill the full MUSE FoV. We used the dedicated sky observations to build the spectra consisting of sky lines and sky continuum. We applied the sky subtraction method ``model'' within the scientific post-processing pipeline stage, which re-fits the sky lines before subtracting them from the on-target data. The sky continuum is subtracted directly without further modifications. Sky spectra were always associated with the closest in time on-target exposures. Such sky subtraction worked well in the blue part of the spectra ($<700$ nm) and is sufficient for this study. Nevertheless, the process was less successful in the red part, and we point the reader to \citet{2024MNRAS.530.3278D} for further details.

The flux calibration was calculated using the ESO-approved MUSE standard stars, which were in almost all cases taken at the same night as the M3G observations, and reduced with the MUSE pipeline recipes. As we did not observe telluric stars, our initial \citep{2018MNRAS.477.5327K} telluric correction was based on the standard stars and was not effective. We improved the telluric correction by using the \textsc{molecfit} software \citep[v1.5.9;][]{2015A&A...576A..77S}, which computes a theoretical absorption model based on radiative transfer and atmospheric molecular line database. We first ran the scientific post-processing part of the pipeline with telluric correction switched off, preparing data cubes with spectra containing the telluric absorption features. We selected spectra of foreground stars, as featureless as possible, as inputs for \textsc{molecfit}. The spectra were taken from merged data cubes combining all on-target exposures (see below). A few galaxies did not have suitable stars, and we used instead the central spectrum of the galaxy as the input for \textsc{molecfit} to calculate the telluric absorption. These telluric spectra were then applied as a correction to all spectra in the merged data cubes. As shown in \citet{2021A&A...649A..63P} and \citet{2024MNRAS.530.3278D}, this procedure significantly improves the spectra and enables the analysis of emission lines and chemical properties of M3G galaxies. In this work, we still track the location of the strong sky and telluric features and mask them during the extraction of stellar kinematics. 

The number of exposures per galaxy that were deemed of sufficient quality to be combined is listed in Table~\ref{t:obs}. We created ``white'' light images from each data cube by summing the cube along the wavelength dimension, and used them to provide the relative spatial offsets between the exposures. These offsets (and relative flux weights) were passed to the pipeline recipe for merging (\textsc{muse\_exp\_combine}), which was also used to reject the cosmic rays, applying the ``median'' rejection method. For PGC\,047202, where we combined 101 images, over the area of almost $2^{\prime\prime}\times2^{\prime\prime}$, the merging required a somewhat different approach. We determined the offsets between the individual exposures as before and defined a World Coordinate System with the location of the four quadrants of the mosaic. Then we used \textsc{muse\_exp\_combine} to combine all 101 exposures, but split into seven wavelength regions, each 65 nm long. This resulted in seven combined data cubes with short wavelength ranges but full spatial extent. Those were further combined into a final cube covering the full spectral range, using \textsc{muse\_cube\_combine}\footnote{\textsc{muse\_cube\_concatenate} could also be used.}.

We note that until this point, all calibrations and corrections (including the sky subtraction) were done on the so-called MUSE Pixel Tables. The product of the merging routines is a combined data cube, which we use from this point for further analysis. The telluric correction was, however, not done on the pixel tables, but on the merged (non-telluric-subtracted) cubes. The final data cubes of the M3G sample are oriented with North up and East to the left, have $0.2^{\prime\prime}\times0.2^{\prime\prime}$ spatial sampling, and a spectral sampling of 1.25 \AA. 

\section{Stellar kinematics extraction}
\label{s:kin}

M3G galaxies are in dense environments and often surrounded by smaller satellites, which partially obstruct the view of the main galaxy. We have mitigated the impact by performing several different extractions of the stellar kinematics via the penalised Pixel Fitting method\footnote{\url{https://pypi.org/project/ppxf/}} \citep[pPXF;][]{2004PASP..116..138C, 2017MNRAS.466..798C}. These included a pPXF fit to a global spectrum, extracted within an elliptical ``effective'' aperture defined by the $R_e$ and the galaxy flattening (Table~\ref{t:m3g}). This was followed first by a single component fit, and then by a multiple component fit to each spectrum, with two different parametrisation of the line-of-sight velocity distribution (LOSVD). The extraction presented here is broadly similar to that of \citet{2018MNRAS.477.5327K}, but it differs in several key points, superseding the previously published kinematics. Nevertheless, the main result of \citet{2018MNRAS.477.5327K}, pertaining to the distribution of the kinematic misalignment angles, remains valid. 

\subsection{Spatial binning and pPXF set up}
\label{ss:bin}

Starting from the final telluric-corrected data cubes, we proceeded to spatially bin them to obtain a uniform signal-to-noise ratio (S/N) across the FoV. We first estimated the S/N for each spectrum in the cube using regions of the spectra lacking emission lines, sky or telluric residuals and strong absorption lines, with the noise estimated from the pipeline-produced variance spectra. Contrary to the previous approach, we only spatially masked prominent stars or compact sources, but included the satellite galaxies. Stars and compact objects to be masked were automatically detected via the subtraction of a smooth MGE model of the "white light'' images: associated masks were built using a $1-\sigma$ threshold above the median residual value. We also excluded all spectra that had an estimated S/N $<2$: those are located at the edges of the MUSE FoV and are dominated by sky subtraction residuals. We used the Voronoi binning method\footnote{\url{https://pypi.org/project/vorbin/}} \citep{2003MNRAS.342..345C}, and set the target S/N to 50. 

For galaxies that have no satellites, the pPXF setup was relatively standard: we parameterised the LOSVD with Gauss-Hermite moments \citep{1993MNRAS.265..213G, 1993ApJ...407..525V}, with the mean velocity $V$, the velocity dispersion $\sigma$ describing the Gaussian part, and the Hermite coefficients $h_3$ and $h_4$, describing the asymmetric and symmetric departures from the Gaussian, respectively. We used additive polynomials of the 12th order, masked all emission lines and a few spectral regions likely to have sky or telluric residuals. Before fitting, the spectra were de-redshifted using an estimated redshift (equivalent to typically $\approx20$ nm for galaxies in our sample), and the fit was constrained between 470 -- 680~nm (rest frame).  We used the MILES stellar library \citep{2006MNRAS.371..703S, 2011A&A...532A..95F} as stellar templates, convolved to follow the MUSE line-spread function \citet[LSF;][]{2017A&A...608A...5G}. As a first step, we fitted the sum of all spectra within the elliptical effective aperture with the full set of MILES stellar templates, producing an {\it optimal} template. This optimal template was used in the subsequent pPXF run to fit all individual Voronoi bins, considerably speeding the fitting process and error estimation. We performed a test by generation optimal templates for each bin, but the resulting kinematics was not different compared to the nominal. Furthermore, as a test we extracted kinematics of a few galaxies using the higher spectral resolution X-shooter Stellar Library \citep{2022A&A...660A..34V}, which produced essentially identical kinematics and we opted to keep using the MILES templates. Finally, we tested extraction using the Calcium Triplet complex (typically redshifted to about 900 nm), but the extracted kinematics did not improve compared to the nominal, and the velocity moments had larger errors due to the increase noise caused by the residuals of the telluric correction.

   \begin{figure*}[h]
    \includegraphics[width=\textwidth]{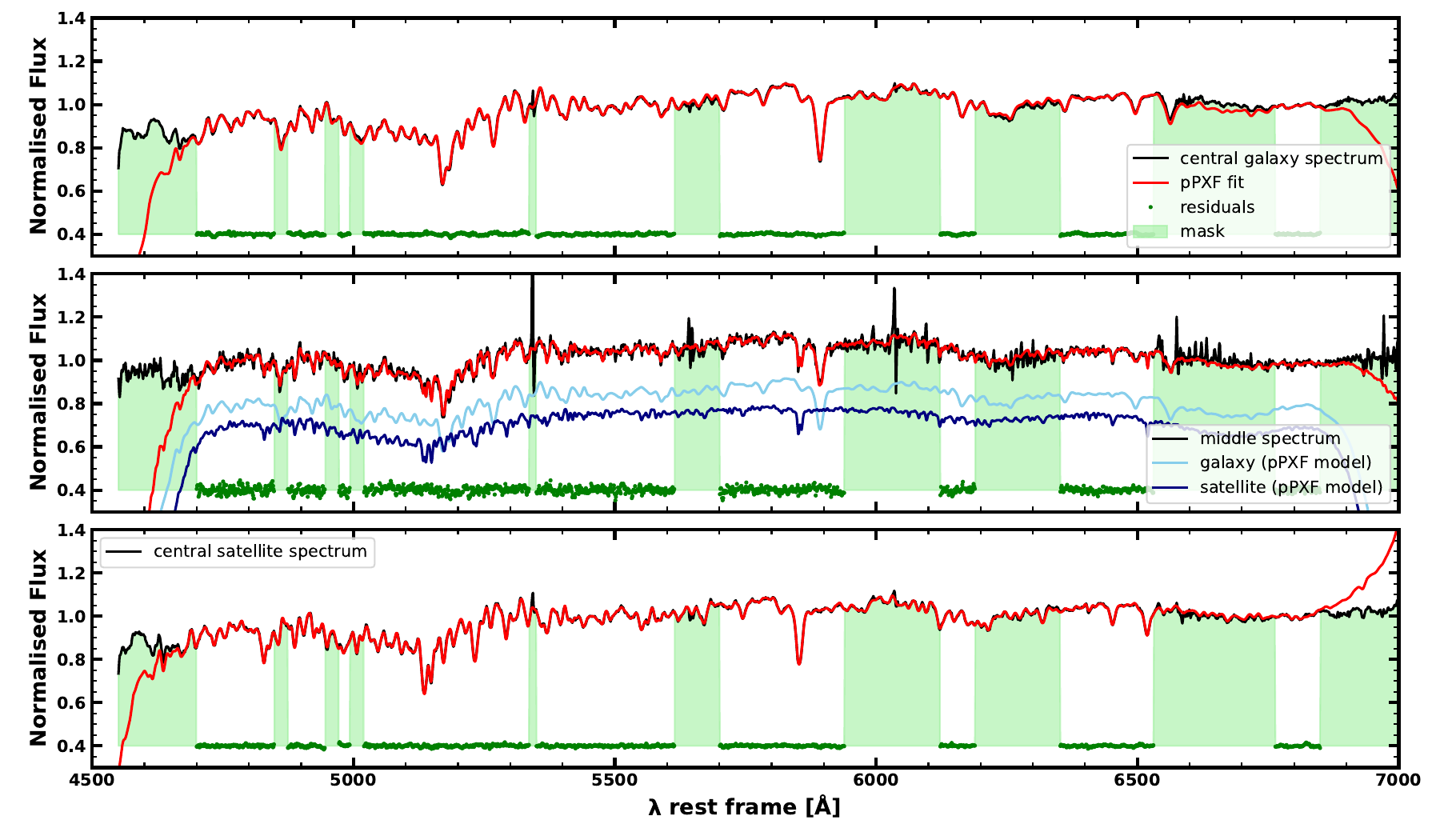} 
      \caption{Spectral fits using pPXF for three different spectra from the PGC\,099188* MUSE field, which features a large interloping satellite. Panels from top to bottom show: the central spectrum of PGC\,099188*, a spectrum from the midpoint between the centre of PGC\,099188* and the interloping satellite, and the central spectrum of the interloping satellite. In each panel the black line shows the observed spectrum, the red line the pPXF fit, with the residual of the fit in green points at the bottom. Green shaded polygons are masked regions excluded from the fit. The middle panel has two additional spectra shown by light and dark blue colours, belonging to the main galaxy and the satellite, respectively. These are model spectra vertically shifted for an arbitrary amount, obtained from the pPXF fits highlighting the satellite and the galaxy contributions to the total spectrum.}
         \label{f:ppxf_exmp}
   \end{figure*}

The presence of overlapping satellites made the extraction of the stellar kinematics quite challenging, as some bins could present a mixed contribution of both the main target and a satellite. We took advantage of the different systemic velocities of those two contributions to implement a multi-component fitting process. Our approach is similar to the one commonly used for separating kinematic components within a given target \citep[e.g][]{2011MNRAS.412L.113C, 2013MNRAS.428.1296J, 2018MNRAS.477.1958C}, except for the fact that we do not use any priors on stellar populations or light distributions \citep[e.g.][]{2012MNRAS.422.2590J, 2019A&A...628A..91S}. When fitting multiple components, we kept the same pPXF set-up as outlined above, and allowed for one ``main'' component with a Gauss-Hermite LOSVD and other components (satellite) described by a single Gaussian LOSVD. Optimal templates were built separately for the central object and the satellites using high S/N regions for each of those. These fits further provide the starting kinematic values ($V$ and $\sigma$) that will be used as a constraint for the subsequent multiple-component runs.

Once both the single and multiple components fits were done for each spatial bin of the MUSE FoV, we looked at the reduced $\chi^2$ statistics of the pPXF results to determine which spatial bin belongs to the main galaxy (i.e.\,dominated by its light), to an intruding satellite, or its spectra could be decomposed into two components. We did this by imposing $\sigma_{sat}\geq \sigma_{inst}\approx50$ km/s in each bin and the following restrictions on $\chi^2_{\rm R} = \chi^2_{\rm single}/\chi^2_{\rm multi}$, the ratio of the $\chi^2$ obtained from the single and multi-component pPXF fits. Bins with $\chi^2_{\rm R}<1.03$ we assigned solely to the main galaxy. Those with $1.03 < \chi^2_{\rm R}<1.05$ were deemed to carry both contributions from the galaxy and a satellite which could be disentangled. Our tests showed that bins with $\chi^2_{\rm R}>1.05$ provide unreliable kinematics for the main galaxy. These bins are dominated by the satellite's light, their kinematics are typically robust, and we assigned them to the satellite. 

In Fig.~\ref{f:ppxf_exmp} we show an example pPXF fit to three different spectra coming from PGC\,099188*, which has a large interloper satellite galaxy in the north east region. The top and bottom panel spectra belong to the main galaxy and the satellite, respectively, while the middle panel spectrum is taken from a location approximately midway between the centres of the galaxy and the satellite. The spectral features in the middle panel clearly indicate that the spectrum is composed of two components, which can be separated into a galaxy spectrum and a satellite spectrum. The interloping satellite is approximately 2000 km/s ahead of PGC\,099188*, which is quite similar to some other cases in the sample. The main galaxy features are, however, not visible in the spectrum taken from the central region of the satellite. Nevertheless, the multi-component pPXF fit can recover the kinematics of the sample galaxy, which is of interest for our study, over a large spatial range. The middle panels of Fig. S2 in the Supplementary Material show the results of the multi-component fit in the case of PGC\,099188, and the separation of the kinematics of the main galaxy and the satellite. More details about the separation of the satellite kinematics are provided in Appendix~\ref{app:sat}.

Preliminary mean velocity maps of the M3G galaxies were presented in \citet{2018MNRAS.477.5327K}. Here we present the improved mean velocity maps, together with maps of velocity dispersions and $h_3$ and $h_4$ Gauss-Hermite moments. All 25 galaxies can be seen in Fig.~\ref{f:kinem}. In a few cases, there are notable differences between the new and old velocity maps, mostly in the regions covered by the outer bins and typically beyond $1 R_e$: those are mostly driven by an improved accounting for the zeroth order velocity term (Section~\ref{sss:var_sys}). We note the measurements of the global velocity dispersions $\sigma_e$, presented in Table~\ref{t:m3g}, were done by summing all spectra within a half-light aperture and parametrising the LOSVD with a Gaussian distribution, applying the single-component pPXF set-up.

\section{Kinematic analysis}
\label{s:kina}

In this section, we analyse different aspects of the extracted M3G kinematics. We start with the ``kinemetric'' analysis to describe and classify the features on velocity maps. We measure the spin parameter as a proxy for stellar angular momentum and classify the M3G galaxies into slow and fast rotators. Finally, we look into the properties of the third Gauss-Hermite moment and investigate the origin of regular rotation in massive galaxies.

\subsection{Kinemetric analysis of velocity maps}
\label{ss:kinem}

We apply the \textsc{kinemetry}  method \citep{2006MNRAS.366..787K} stepping through a velocity map at different radii, and defining an ellipse at each radius along which the azimuthal variation of velocities are best reproduced by a simple cosine $V (R)= V_{\rm rot}(R)\times\cos(\theta)$. The optimisation is done via harmonic analysis, following 
\begin{equation}
\label{eq:kin}
V(R, \theta) = A_0(R) + \sum_{n=1}^{N} A_n(R)\sin(n\theta) +\sum_{n=1}^{N} B_n(R)\cos(n\theta).
\end{equation}
\noindent For odd moments, such as the mean velocity map, $B_1 (R) = V_{\rm rot}(R) = V_c\sin(i)$. Therefore, harmonic terms up to $N=3$, with the exception of $B_1$, are fitted to define the parameters of the ellipse, while the $N\ge5$ terms are not fitted, but calculated once the best-fitting ellipse is defined. They represent the deviation from the assumed cosine-law azimuthal velocity variations. $A_0$ is a constant term, which can be subtracted from each ellipse ring, revealing the relative motions within the galaxy.

When analysing maps of stellar velocities, it is useful to define amplitude terms as $k_n = \sqrt{A_n^2 + B_n^2}$ \citep{2008MNRAS.390...93K}. $k_1$ is dominated by $B_1$ and describes the rotation, while $k_5$, based on the first pair of harmonic terms that are not minimised, is typically used to estimate the deviation from the cosine-law.  \citet{2011MNRAS.414.2923K} required that a unit-less quantity $\overline{k_5/k_1}\le0.04$ for a velocity map to be well described by the cosine-law, where the average is within the effective radius. Such maps are characterised as having {\it regular rotation}, while maps with $\overline{k_5/k_1}>0.04$ are described as having {\it non-regular rotation}. We also note that in this notation $k_0 = A_0$.

We ran \textsc{kinemetry} in several setups and on two types of maps tracing the LOSVD moments: on the flux images \citep[performing classical isophotometric analysis;][]{1985ApJS...57..473L,1987MNRAS.226..747J} and on the mean velocity maps.  The analysis of the velocity maps were separated into two steps: the first one dealing with the removal of the zeroth kinemetry term (related to the systemic velocity), and the second with the actual harmonic analysis of the corrected velocity maps.

\subsubsection{Spatial variations of the $k_0$ term}
\label{sss:var_sys}

The outer regions of some velocity maps extracted in Section~\ref{s:kin} display an unusual symmetric distribution of velocities with respect to the principal axes. Two examples are presented in the first column of Fig.~\ref{f:vsys_rem}. Velocity maps of massive galaxies are expected to exhibit no rotation across the field ({\it Group a}), irregular and low amplitude rotation ({Group b}), or spatially localised rotating components exhibiting point-antisymmetric velocities fields ({\it Group c}). The velocity maps in Fig.~\ref{f:vsys_rem}, however, have both non-zero velocities and are {\it symmetric} at large radii. \textsc{Kinemetry} analysis indicates that in these galaxies the $k_0$ term is changing with radius, independently from the higher terms, notably $k_1$, which traces the rotation around the galaxy centre.

A similar symmetric change in velocities was also noticed by \citet{2015ApJ...807...56B} in NGC\,6166, the BCG of Abell\,2199, where the velocities at larger radii approach the average velocity of the cluster galaxies. A possible explanation is that the halo of the BCG consists of tidal debris, which is more at rest with respect to the cluster than the central galaxy. If we combine the systemic velocities\footnote{As we de-redshifted the spectra before extracting the kinematics, the systemic velocity is estimated from the redshift of the galaxy using Eq.~(9) of \citet{2017MNRAS.466..798C}. See also \citet{1999astro.ph..5116H}.} of galaxies on Fig.~\ref{f:vsys_rem} with their $k_0$ values, $V_{\rm sys} +k_0$ approach the values of the respective mean cluster velocities. For example, PGC047202 $V_{\rm sys}$ is 700 km/s below the cluster mean velocity $V_c=14450$ km/s (Table~S2 of the Supplementary Material), still within the spread of the cluster velocities $\sigma_c=1007$ km/s \citep{2014ApJ...797...82L}. At the edge of the MUSE field, even though only at $2 R_e$ and well within the main galaxy, there is a significant decrease in the difference to $\sim590$ km/s. 

\begin{figure}
    \includegraphics[width=\columnwidth]{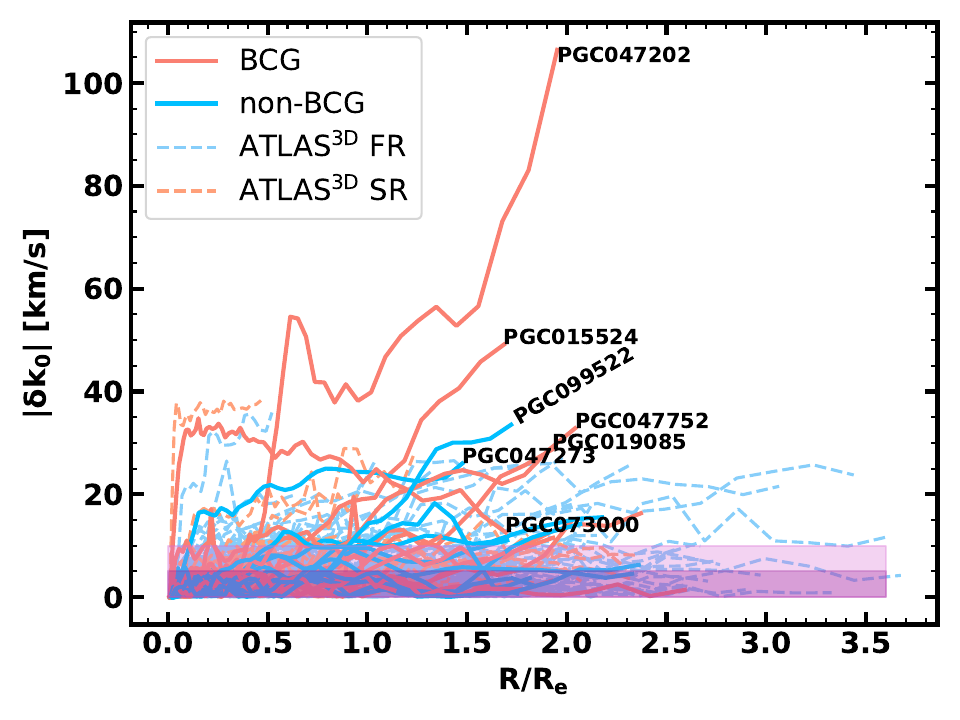} 
      \caption{The radial profiles of the absolute $k_0$ variation for M3G galaxies ($\delta k_0 = |k_0(0) - k_0(R)|$), compared to the same from the ATLAS$^{\rm 3D}$ galaxies. M3G galaxies are divided in BCGs (red lines) and non-BCGs (blue lines), while ATLAS$^{\rm 3D}$ galaxies are divided into fast (light blue dashed lines) and slow (light red dashed lines) rotators, respectively. Purple shaded regions show the standard deviation of all ATLAS$^{\rm 3D}$ and M3G (darker shaded regions) velocity errors. Names highlight specific galaxies with the largest difference between the maximum and the minimum values of $k_0$: $\Delta k_0>20$ km/s. }
         \label{f:vsys}
   \end{figure}

The majority of M3G galaxies has nearly constant $k_0$ across the field. We quantify the absolute radial change as $\delta k_0 = |k_0(0) - k_0(R)|$, and introduce the maximum change $\Delta k_0  = \max(k_0(R)) - \min(k_0(R))$ as the difference between the maximum and the minimum $k_0$ values\footnote{Note that it is possible that $\max(\delta k_0)<\Delta k_0$, if $\min(k_0(R))<k_0(0)$, but this is rare and within a few km/s, except for PGC\,047177 and PGC\,047202, where the differences are 11 and 17 km/s, respectively. }. We show the radial change $\delta k_0$ in Fig.~\ref{f:vsys} and list $\Delta k_0$ in Table~\ref{t:m3g}. There are seven galaxies that have $\Delta k_0 >20$ km/s (28\%), five of which are BCGs and two are non-BCGs. 

We compare in Figure~\ref{f:vsys} $\delta k_0$ with those from the ATLAS$^{\rm 3D}$ sample from \citet{2011MNRAS.414.2923K}. The M3G galaxies are separated into BCGs and non-BCGs, while the ATLAS$^{\rm 3D}$ are separated into fast and slow rotators. The majority of galaxies in both samples do not show large variations in $\delta k_0$. Only 19 ATLAS$^{\rm 3D}$ are found to have $\Delta k_0 >20$ km/s (7\% of the sample). 

In the ATLAS$^{\rm 3D}$ sample notable outliers are NGC\,4486 and NGC\,4753 (in order of decreasing $\Delta k_0$). NGC\,4486 is a massive slow rotator with an active SMBH and complex stellar kinematics \citep{2014MNRAS.445L..79E}. $\delta k_0$ is relatively constant at $\sim$35 km/s, except for a sudden jump between the central value and the values at radii larger than 2\arcsec. The nucleus of NGC4486 is dominated by AGN activity \citep[for examples of central spectra of NGC4486 at different spatial resolutions and the influence of the AGN emission see][]{2024MNRAS.527.2341S}. Given that this type of $\delta k_0$ profiles are rare, the issue is likely related to difficulties in extracting reliable stellar kinematics from the nucleus. On the other hand, NGC\,4753 is a galaxy with disturbed large scale morphology, indicating a recent merger \citep{2020MNRAS.498.2138B}, and its $\delta k_0$ changes gradually with radius. NGC\,4486 and NGC\,4753 could be taken as showcases for two different reasons of $k_0$ radial variations: due to difficulties in extracting the correct nuclear velocities\footnote{See also \citet{2021ApJ...909..141P} for a case of a moving supermassive black hole as a potential cause of the $k_0$ variation within the nucleus.} and due to disturbed equilibrium at large radii.

Fig.~\ref{f:vsys} highlights seven M3G galaxies of which the largest $\delta k_0$ variation is seen for PGC\,047202, the brightest and the largest galaxy in the sample, also found in the densest environment. A counter-example is PGC\,073000, characterised with a sudden jump within the central $2\arcsec$ reaching a value similar to NGC4486, which stays constant within $\sim$$10\arcsec$\, and then steadily declines. The central sudden variation could be related to the contamination of the nuclear spectra, as this galaxy has very complex gas kinematics, ionised by an AGN \citep{2021A&A...649A..63P}. PGC\,015524, with the second highest change in  $V_{\rm sys}$, has a similar behaviour to PGC\,047202, with a steady increase of $\delta k_0$ from the central value. The increase is negligible (less than the typical velocity uncertainty) in the central $15\arcsec$, but afterwards increases rapidly. PGC\,047273 is a non-BCG galaxy, and the only one with regular kinematics (see Section~\ref{sss:nonR}), with a significantly varying $\delta k_0$, which also shows a jump from the central velocity within $2\arcsec$. This galaxy also has a nuclear dust disk, unusually located in a near polar orientation, with LINER ionisation properties \citep{2021A&A...649A..63P}. Finally, PGC\,047752, PGC\,019085 and PGC\,099522 are also worth highlighting, with $\Delta k_0>20$ km/s beyond the effective radius. PGC\,099522 is a non-BCG, but its $\delta k_0$ is obtained from only half of the field as its velocity map is obstructed by two satellite galaxies (see Appendix~\ref{app:sat}). The extraction of \textsc{kinemetry} parameters are therefore less secure and $\delta k_0$ is uncertain, and we will not consider it as an $k_0$ outlier. In summary, two galaxies have a central variation of $k_0$ (PGC\,047273 and PGC\,073000), and four galaxies have variable $k_0$ beyond the effective radii, where PGC\,047202 shows the strongest variation from about $0.5\times R_e$ outwards. 

   \begin{figure*}
    \includegraphics[width=\textwidth]{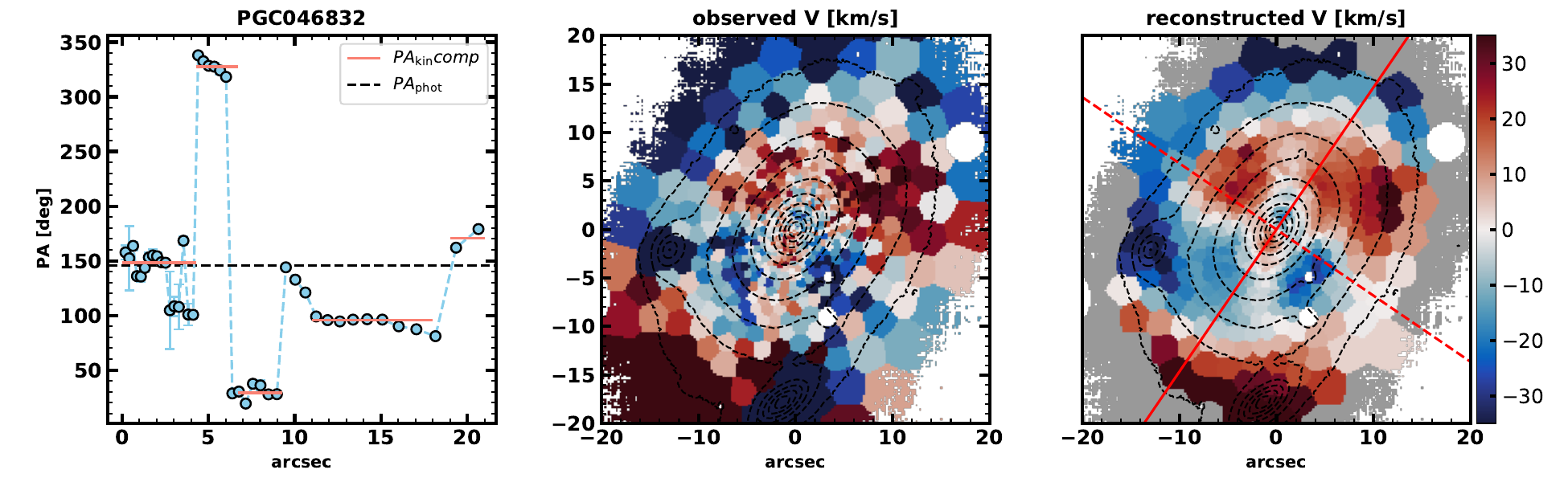} 
      \caption{An example of a multi-spin galaxy: PGC\,046832, which has five visible changes of the rotational orientation. {\it Left}: kinematic positions angle as a function of radius (1 kpc $\sim$ 1\arcsec), with the photometric major axis shown in dashed black line and median position angles of five components highlighted with short orange lines. {\it Middle}: observed velocity map. {\it Right}: reconstructed velocity map using low-order \textsc{kinemetry} coefficients, highlighting only the rotational component of the velocity map. Red solid and dashed lines are the project major and minor axes of the galaxy, respectively. Along the major axis in projection, there are three reversals of the short-axis rotation (prograde, retrograde, prograde), while along the minor axis in projection, there are two (prograde and retrograde), making this galaxy a case of five multi-spin components.}
         \label{f:ms} 
   \end{figure*}

Separating the ATLAS$^{\rm 3D}$ sample into slow and fast rotators shows a potential trend where large variations in the systemic velocity are somewhat more likely for slow rotators (11\% slow rotators compared to 7\% fast rotators). In the M3G sample, excluding the central $k_0$ variation, all galaxies with large $\Delta k_0$ are BCGs and slow rotators (see Section~\ref{ss:am}). 

In Appendix~\ref{app:vsys}, we describe three methods to correct M3G velocity maps for the variation of the $k_0$ term and determine that this variation can be linked with the existence of an additional stellar component. Once the spectral contribution of this components is removed, the new $\delta k_0$ is essentially constant.  Therefore, we conclude that being massive, a BCG, and a slow rotator {\it favours} conditions for existence of multiple kinematics components in the outer regions, likely not in equilibrium with the host galaxy. We will further discuss this issue in Section~\ref{ss:mcc}. Corrected velocity maps are shown with the rest of the kinematics in Fig.~\ref{f:kinem}.

\subsubsection{Non-regularity of M3G velocity maps } 
\label{sss:nonR}

After removing the spatially varying $k_0$ as described in App.~\ref{app:vsys}, we run \textsc{kinemetry} on the velocity maps presented in Fig.~\ref{f:kinem}. We verified that there are no significant differences between \textsc{kinemetry} results obtained on the original or ``cleaned" velocity maps, as one would expect due to the orthogonality of the harmonic coefficients. Minor differences are occasionally found in degenerate parameters that define the shape and orientation of the best-fitting ellipses, but they are fully within the uncertainties. 

All \textsc{kinemetry} parameters were left to vary freely, except the centre, which was fixed to the peak of the surface brightness. In several cases, especially those with very complex velocity maps, the flattening parameter was found not to be  well constrained and oscillate over the field, which happens when the assumption on the cosine law is not justified. To constrain the analysis we restricted the ellipse flattening to be equal or larger than the global flattening of the stellar distribution, $Q_{\rm min}\geq Q_p$. Example plots of kinemetric parameters ($PA_{\rm kin}$, $Q_{\rm kin}$, $k_1$ and $k_1/k_5$) for all galaxies are presented in Fig. S3 of the Supplementary Material. Relevant values are presented in Table S2. 

The primary use of \textsc{kinemetry} is to quantify irregularities on velocity maps and to highlight kinematic features. We follow the definitions from \citet{2011MNRAS.414.2923K}, where the {\it Regular Rotators} (RR) are galaxies with $\overline{k_5/k_1}\le0.04$, while {\it Non-Regular Rotators} (NRR) have $\overline{k_5/k_1}>0.04$. The average values are weighted by luminosity and obtained using eq.~(1) from \citet{2008MNRAS.390...93K}, see also \citet{1999ApJ...517..650R}.

We applied the above-mentioned classification into kinematic groups a) to e), with a small modification for galaxies that exhibit multiple velocity reversals; they are flagged as multi-spin (MS)\footnote{\citet{1994AJ....108..456R} was first to use the ``multi-spin'' term to define galaxies ``with more than one axis of rotation''.}. Therefore, a  galaxy showing two spin reversals is of type MS-2 \citep[e.g. a counter-rotating KDC such as IC\,1459, {\it Group c},][]{1988ApJ...327L..55F}. Furthermore, when the difference between the $PA_{\rm kin}$ and the photometric major axis ($PA_{\rm phot}$) is close to $90\degr$, the rotation is {\it around} the long-axis. If such a galaxy has two spin reversals, where one components is in a long-axis rotation, it will be classified as MS-2L, ``L'' standing for  {\it long-axis rotation} \citep[e.g. the classic KDC NGC\,4365,][]{2001ApJ...548L..33D}. For distinction, we will refer to the much more common rotation around the minor axis as {\it short-axis rotation}. This nomenclature is based on the connection of the orbital families and the type of rotation they generate: the short-axis tubes rotate around the short (minor) axis of a galaxy, while the long-axis tubes rotate around the long (major) axis of a galaxy\footnote{We note that there were various ways to name the rotation around the long-axis in the literature, including ``prolate rotation'' or ``minor-axis rotation''  \citep[e.g.][]{1988A&A...195L...5W}.}. 

Velocity maps of eight M3G galaxies have two spin reversals (KDCs), and there are at least four galaxies with evidence for more than two kinematic components oriented in different directions. PGC\,046832 (MS-5L) is perhaps the most complex example, and we show it in Fig.~\ref{f:ms}. Looking along the major axis of the galaxy (traced by the solid line on the rightmost panel), one can see three spin reversals, the first within central 3\arcsec, the second between $\sim5-10\arcsec$ and the third beyond 10\arcsec. While these flips between the prograde and retrograde motions are seen in projects, we nominally associate them with to the short-axis rotation. Only detailed modelling \citep[as done in ][]{2021MNRAS.508.4786D} can associate visible structures with internal orbital families. Looking along the minor axis (traced by the dashed line) there are two spin reversals, the first $\sim3-8\arcsec$ and the second beyond 10\arcsec. We will associate these with the long-axis rotation reversals. While the orientation of each spin direction is not exactly aligned to the projected major and minor axes, one can count a total of five distinct spin reversals. 

Thirteen galaxies, or about half of the M3G sample, have only one kinematic component (i.e. MS-1). Two of these galaxies have long-axis rotation over the full map, while five galaxies can be classified as RR, with short-axis rotations and nearly constant $PA_{\rm kin}$. The velocity maps of other MS-1 galaxies are characterised with kinematic twists (KT) with a steady progression of the $PA_{\rm kin}$ (both converging to and diverging from the photometric major axis). Long-axis rotation is often found in the innermost component (e.g. PGC\,047752, MS-4L), but other options are possible: PGC\,047590 (MS-3L) has two prolate-like components that are also counter-rotating with respect to each other, or PGC\,007748 (MS-2L) with an outer long-axis rotation. The main characteristic of galaxies with multiple spin components is that all, with a possible exclusion of PGC\,046785, have at least one component in (approximate) long-axis rotation. The central component of PGC\,046785 is misaligned by about $50^{\rm o}$ from the photometric major axis, sufficiently distinct from rotation around the major axis, but strongly misaligned. 

The existence of RR galaxies in the M3G sample is quite striking, considering that they are all very massive galaxies. They are, however, only found in non-BCGs, typically fainter galaxies in our sample.  In Abell\,3558, the brightest RR galaxy (PGC\,047177) is ranked as the 4th brightest galaxy in the cluster. In Abell\,3556, the second brightest galaxy (PGC\,046785) is marginally a RR, with $\overline{k_5/k_1}=0.037\pm0.004$ within the effective radius, but becomes a NRR at larger scales. There are no RR in Abell\,3562, where the luminosity differences between the BCG and other three members are the smallest.

In summary, massive galaxies often have complex stellar kinematics with multiple spin reversals, while the long-axis rotation is common. Nevertheless, some exhibit regular velocity maps consistent with disk-like rotation, but these are typically less bright (massive) galaxies.

\subsection{Stellar angular momentum}
\label{ss:am}

\begin{figure}
    \includegraphics[width=\columnwidth]{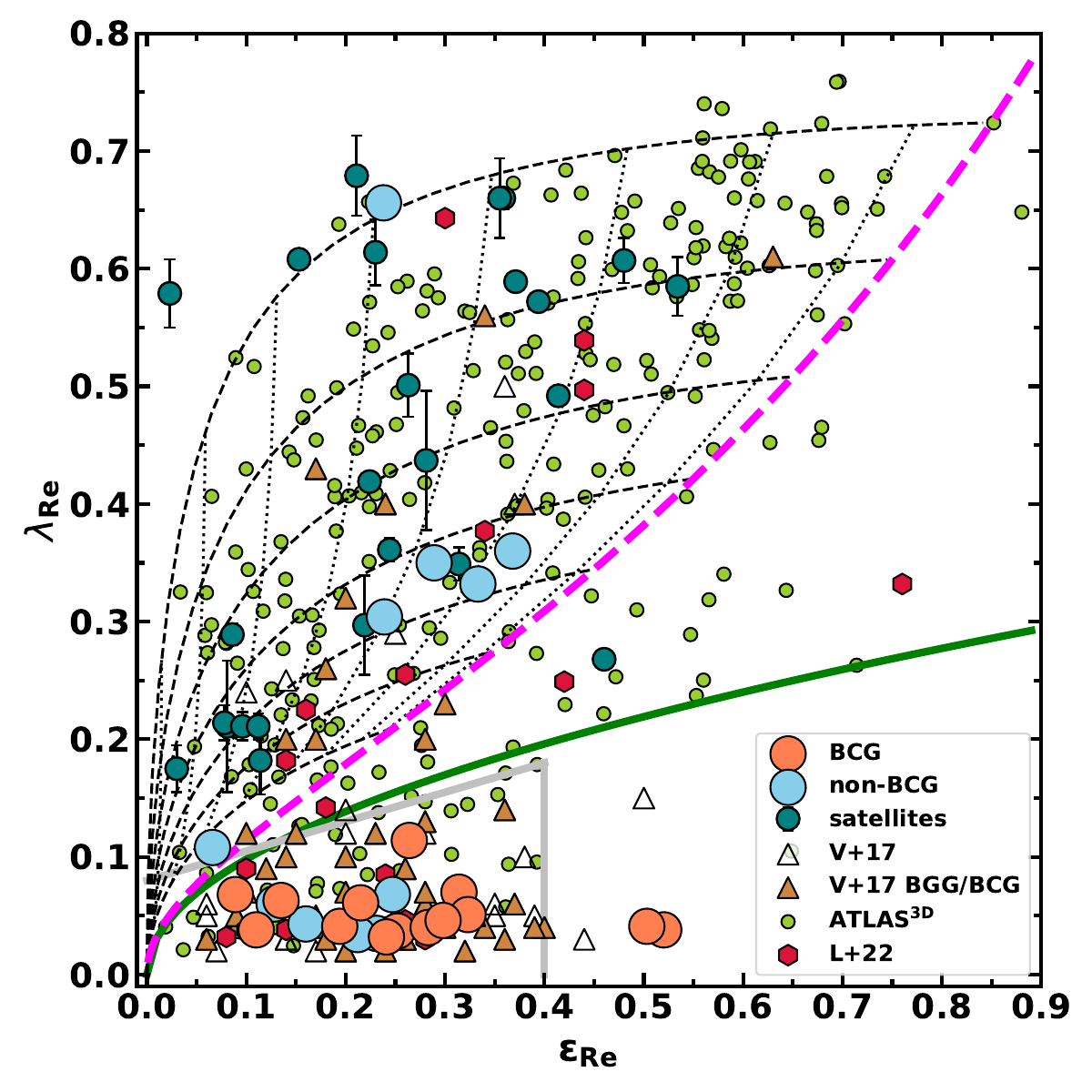} 
      \caption{$\lambda_{\rm Re}$ versus ellipticity $\epsilon_{\rm Re}$ for the M3G sample and galaxies drawn from several reference surveys. M3G galaxies are shown as large red (BCGs) and blue (non-BCGs) circles, with uncertainties smaller than the symbols. Small satellites of M3G galaxies found in the MUSE FoV are drawn as teal circles, while small light green circles are ATLAS$^{\rm 3D}$ galaxies. Triangles are galaxies from the MASSIVE survey, where filled (brown) triangles are their BCG or brightest group galaxies (BGG), taken from \citet{2017MNRAS.464..356V}. Dark red hexagons are BGGs from \citet{2022MNRAS.515.1104L}. Dark green solid line separates fast from slow rotators as $0.31\times \sqrt{\epsilon_{\rm Re}}$ \citep{2011MNRAS.414..888E}, while the grey solid line, given by \citet{2016ARA&A..54..597C} as $0.08+\epsilon/4$, is an alternative separatrix. The dashed magenta line shows the edge-on view for ellipsoidal galaxies integrated up to infinity with $\beta = 0.7\times \epsilon$ \citep{2007MNRAS.379..418C}. The black dotted lines show the location of galaxies, originally on the magenta line, as the inclination is varied by $10\degr$, while the black dashed lines trace locations for galaxies with intrinsic ellipticities from 0.25 (lowest) to 0.85 (highest) in step of 0.1. }
         \label{f:lr} 
   \end{figure}

We follow \citet{2011MNRAS.414..888E} for calculating the global stellar angular momentum $\lambda_{\rm Re}$ within an elliptical aperture of ellipticity $\epsilon_{\rm Re}$, which has the same area as the circle of radius $R_{\rm e}$. The uncertainties were estimated by perturbing the velocity and velocity dispersion maps via Monte Carlo simulations. Figure~\ref{f:lr} shows the $\lambda_{\rm Re} - \epsilon_{\rm Re} $ diagram for the M3G galaxies, plotted together with the results from the ATLAS$^{\rm 3D}$ \citep{2011MNRAS.414..888E} and MASSIVE surveys \citep{2017MNRAS.464..356V}, as well as a selection of nearby brightest group galaxies observed with MUSE \citep{2022MNRAS.515.1104L}. We also add the satellites of M3G galaxies for completeness, which are all, as expected, fast rotators. $\lambda_{\rm Re}$ values of M3G galaxies are listed in Table~\ref{t:m3g} and Table S1 of the Supplementary Material.  

There are six fast rotators in the M3G sample, all classified as non-BCGs and RR, which are, as mentioned earlier, lower luminosity galaxies in Abell\,3556 and Abell\,3558. All M3G BCGs and the remaining five non-BCGs (45\%) are classified as slow rotators, which means 76\% of all M3G galaxies are slow rotators. Within the M3G sample, Abell\,3562 in the SSC has the largest number of slow rotators (4). Notably, slow rotators are found in closer proximity to other more massive galaxies. We show this in Fig.~S1 of the Supplementary Material where we plot the M3G SSC subsample over the distribution of galaxies with spectroscopic redshifts within the SSC (z=0.048$\pm$0.004) from the \citet{2018MNRAS.481.1055H} catalogue. 

PGC\,046785, marginally a RR within the effective radius, has also the lowest $\lambda_{Re}$ value among fast rotators. It remains a fast rotator using the alternative \citet{2016ARA&A..54..597C} fast versus slow rotators separation. PGC\,046785 has a strong kinematic twist of 60\degr\,and has three visible multi-spin components. Similarly as it shows significant departures from regular rotation beyond the half-light radius, estimating its $\lambda_R$ and $\epsilon$ values at $(2\times R_e)$ would make it a slow rotator.  In contrast, PGC\,047355 has the largest $\lambda_{Re}$ value in the sample, placing it among one of the fastest rotating ETGs. We also note that PGC\,047202, the brightest (most massive) and the largest galaxy in the sample, found in the centre of the richest cluster, also has the largest $\lambda_{Re}$ among M3G slow rotators. Finally, PGC\,019085 and PGC\,048896 are the flattest galaxies in the sample and occupy the sparsely populated region of $\lambda_{\rm Re} - \epsilon_{\rm Re}$ with $\epsilon>0.4$ \citep{2018MNRAS.tmp..522G}. 

\begin{figure}
    \includegraphics[width=\columnwidth]{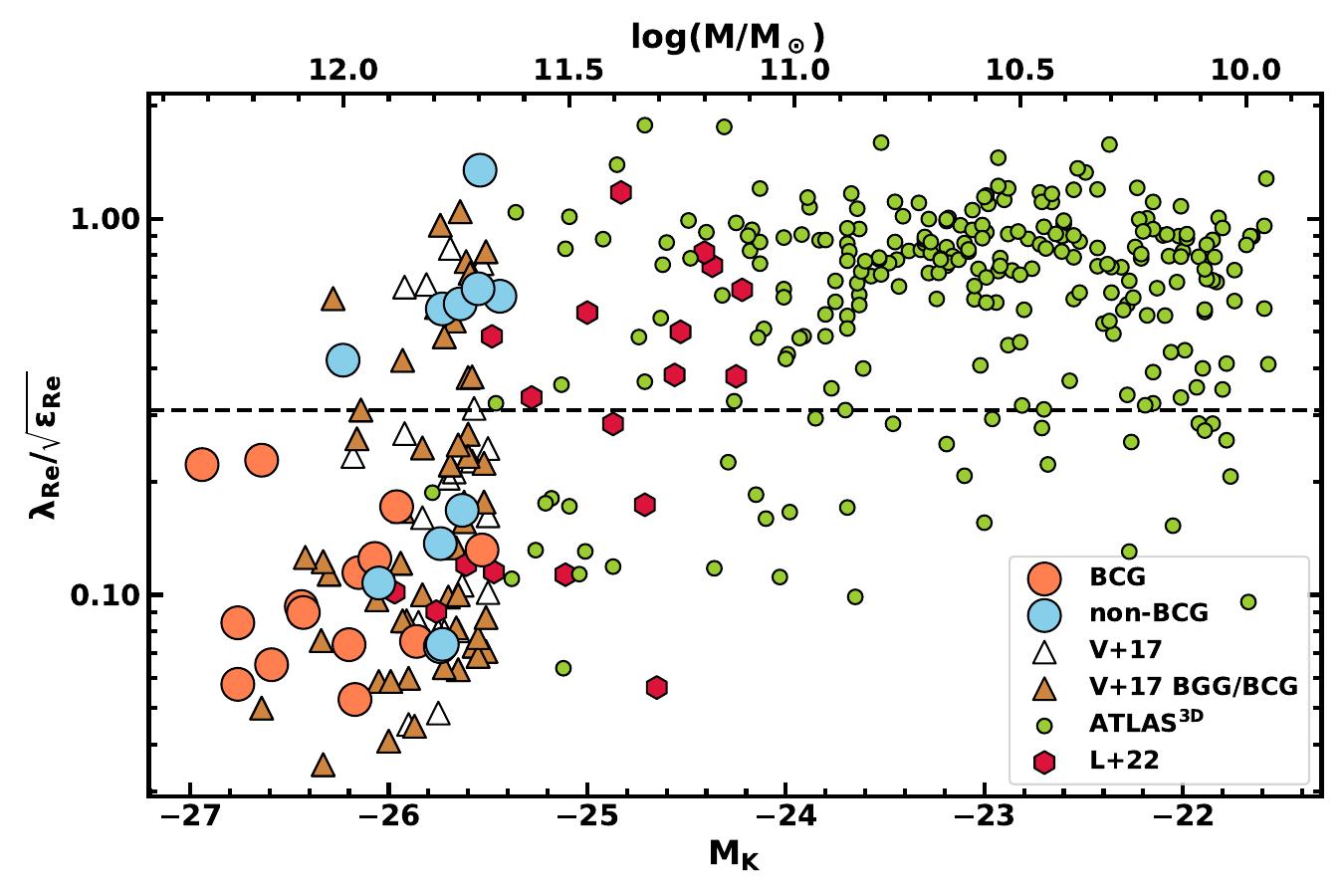} 
      \caption{The distribution of the normalised $\lambda_{Re}/\sqrt{\epsilon}$ spin parameter as a function of the $K_s$-band absolute magnitude. The horizontal dashed line (at 0.31) separates slow (below) from fast (above) rotators. Plotted are galaxies from M3G (large circles), MASSIVE (triangles) and ATLAS$^{\rm 3D}$ (small green circles) samples. M3G galaxies are split into BCG (orange) and non-BCGs (blue), while other galaxies are from the same samples as on Fig.~\ref{f:lr}. The upper axis is in units of log solar masses, obtained with eq. (2) from \citet{2013ApJ...778L...2C}.}
         \label{f:lr_mag} 
   \end{figure}

In Fig.~\ref{f:lr_mag} we show the dependence of $\lambda_{Re}$ on the absolute brightnesses of galaxies. We also use the calibration from Eq.~(2) of \citet{2013ApJ...778L...2C} to transform the $K_s$-band absolute magnitudes into masses, putting all three samples onto the same mass system. Combining the data from the ATLAS$^{\rm 3D}$ and other surveys of massive galaxies \citep{2017MNRAS.464..356V, 2022MNRAS.515.1104L}, one can see a relatively abrupt disappearance of fast rotators above a threshold of $M_K\approx-26$ or $M\approx8\times10^{11}$ $M_\odot$, where there are essentially no fast rotators (exceptions are NGC\,3158 and PGC\,046785).

   \begin{figure*}
    \includegraphics[width=0.4\textwidth]{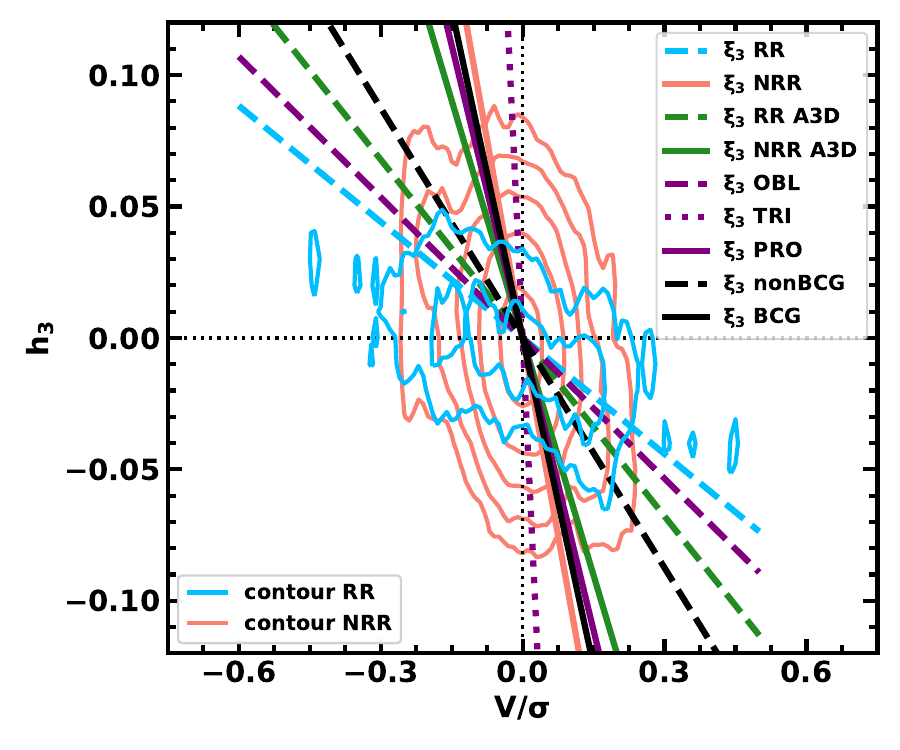} 
    \includegraphics[width=0.6\textwidth]{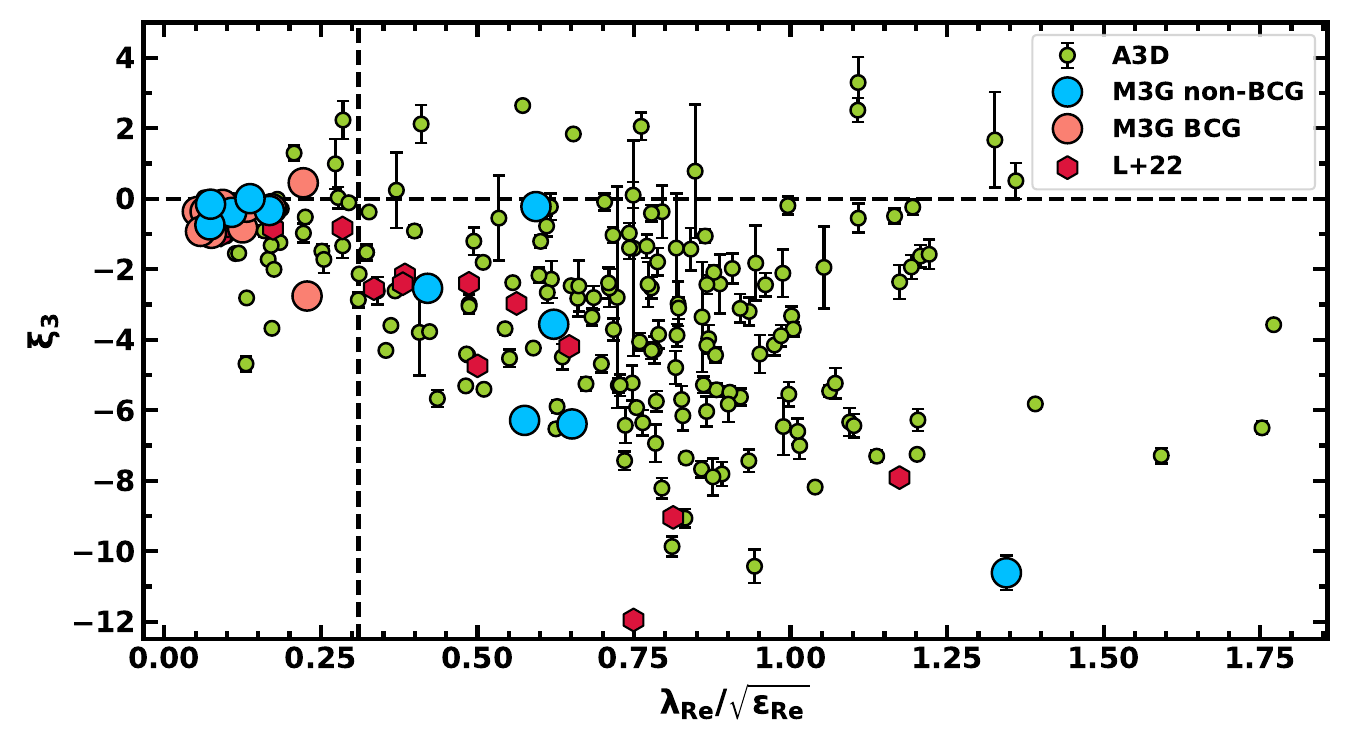} 
      \caption{{\it Left:} Local $V/\sigma - h_3$ relation for every spectrum within the effective radius and with an error on $h_3$ smaller than 0.05 for M3G galaxies. Red contours are for NRR galaxies and blue for RR galaxies, and they are based on logarithmic number counts starting from 0.25 with a step of 0.5. Straight lines show the slopes of $V/\sigma - h_3$ distributions estimated by the $\xi_3$ parameter (see Section~\ref{ss:h3h4} for details) for M3G galaxies separated into RR (blue) and NRR (red), BCGs (black solid) and non-BCGs (black dashed), and three classes based on the kinematic misalignment angle (see Table~S2): ``oblate'' (violet dashed), ``triaxial'' (violet dotted) and ``prolate'' (violet solid). The values for the local ETGs from the ATLAS$^{\rm 3D}$ Survey, also separated into RR and NRR galaxies, are plotted in green. {\it Right:} Distribution of individual $\xi_3$ values as a function of the normalised spin parameter where the value of 0.31 separates slow from fast rotators (vertical dashed line). M3G galaxies are divided into BCGs (red) and non-BCGs (blue) and shown with large red and blue symbols, respectively, while the small light green symbols are for local ETGs from the ATLAS$^{\rm3D}$ survey. Dark red hexagons are from \citet{2022MNRAS.515.1104L}. The error bars on M3G values are typically smaller than the symbols size.}
         \label{f:vs_h3} 
   \end{figure*}

\subsection{Correlation with the $h_3$ Gauss-Hermite moment}
\label{ss:h3h4}

In order to investigate what sustains the relatively high $\lambda_R$ values in some massive galaxies we make use of the Gauss-Hermite $h_3$ moments, which measure skewness or asymmetric deviations of the LOSVD from a Gaussian distribution. Skewed LOSVDs can be constructed by superimposing two kinematic components with Gaussian LOSVDs, but of different mean values and dispersions. \citet{1995A&A...293...20S} showed that a skewed LOSVD, with a large $h_3$ value, is typical of a rotating disk. The relation between $h_3$ and the normalised mean motions, $V/\sigma$, is particularly useful, as fast rotators display a strong anti-correlation in a $V/\sigma - h_3$ diagram, which is absent for slow rotators \citep{1994MNRAS.269..785B, 2011MNRAS.414.2923K, 2017ApJ...835..104V}. 

\citet{2019MNRAS.489.2702F} introduced a quantitative measure for the $V/\sigma - h_3$ (anti-)correlation using the following expression:
\begin{equation}
\label{eq:xi3}
\xi_3 = \frac{\sum_i F_i h_{3,i}(V_i/\sigma_i)}{\sum_i F_i h_{3,i}^2}
\end{equation}
where the local flux $F_i$, the mean velocity $V_i$, the velocity dispersion $\sigma_i$ and the skewness parameter $h_{3,i}$, measured for each spatial bin $i$ within the kinematic maps, provide an estimate of the global slope of the $V/\sigma - h_3$ relation. The $\xi_3$ parameter is tuned such that when $V/\sigma$ and  $h_3$ are (anti-)correlated, $h_3 = (1/\xi_3)V/\sigma$ and $\xi_3$ is the inverse of the slope of the (anti-)correlation. Based on numerical tests \citep{2019MNRAS.489.2702F}, fast rotating galaxies have $\xi_3<-4$, while slow rotators are expected to have values close to 0, or mildly positive values. We calculated the $\xi_3$ values and list them in Table~S2 of the Supplementary Material. 

The left panel of Fig.~\ref{f:vs_h3} shows the distribution of $V/\sigma - h_3$ values for all individual bins which make up the kinematic maps of M3G galaxies. They are divided into those that belong to RR and NRR galaxies (Table~\ref{t:m3g}). The anti-correlation is evident for RR galaxies, while it is barely existent for NRR galaxies, as emphasised by blue dashed and solid orange lines. We also make comparisons between $\xi_3$ slopes calculated by grouping galaxies in different distributions, for RR and NRR, BCGs and non-BCGs, and based on the distribution of kinematic misalignments in the M3G sample \citep[][the classification is repeated in Table~S2 for completeness]{2018MNRAS.477.5327K}. We also add the $\xi_3$ values for the RR and NRR galaxies from the ATLAS$^{\rm 3D}$ sample of local ETGs \citep{2011MNRAS.414.2923K}. The comparison indicates that M3G RR, ``oblate'' (aligned) galaxies, and non-BCGs have more negative values of $\xi_3$, similar to the value given by the local RR ETGs. M3G NRR, BCGs and ``prolate'' (highly misaligned) galaxies have $\xi_3$ values similar to those of local NRR ETGs. The $\xi_3$ values relating to the slopes of various samples in Fig.~\ref{f:vs_h3} are listed in Table S3 in the Supplementary Material.

The comparison of the $\xi_3$ values for the M3G Sample (Table S2), as well as for the ATLAS$^{\rm 3D}$ and the \citet{2022MNRAS.515.1104L} samples, is shown in the left panel of Fig.~\ref{f:vs_h3}. Errors were estimated using Monte Carlo simulations for each kinematic map ($V$, $\sigma$, $h_3$). As expected, BCGs have low $\xi_3$, with one galaxy showing a positive value. Non-BCGs also often have low $\xi_3$ values, including one that was classified as RR and fast rotator (PGC\,047177). This galaxy indeed does not have a coherent $h_3$ map, in spite of a very regular velocity map and a generally high amplitude of rotation (Fig.~\ref{f:kinem}). Other fast rotators in the M3G sample however show a spread of $\xi_3$ values from $\sim-2$ to $-11$. 

Analysis of the numerical simulations showed that the value of $\xi_3$ depends on the orbital compositions of a galaxy \citep{2009ApJ...705..920H, 2019MNRAS.489.2702F}. Strongly negative values of  $\xi_3$ are characteristic for orbital distributions dominated by short-axis tubes, while values close to zero, or even positive, are found in systems with high fractions of long-axis tubes. This is important as it is not only the existence of stellar disks, but the general existence of coherent and dominant streaming around the short axis that results in large negative $\xi_3$ values. Furthermore, it suggests that the M3G galaxies with high angular momentum contain large fraction of short-axis tubes, which in some cases are completely dominating the orbital distribution \citep[e.g. for $\xi_3~-8.4$ the fraction of short-axis tubes is 60\%,][]{2019MNRAS.489.2702F}. 

\section{Discussion}
\label{s:disc}

The main finding of this work is a surprising richness of features on velocity maps of galaxies with masses above $\sim10^{12} M_\odot$. High quality IFU data reveal velocity maps with  multiple-spin reversals, not seen before. On the other hand, regular rotation and fast rotators persist even among very massive galaxies. 

\subsection{Multi-spin nature of velocity maps of massive galaxies}
\label{ss:ms}

The M3G sample comprises galaxies that do not show net rotation ({\it Group a}: 2/25), galaxies with complex irregular velocities ({\it Group b}: 9/25), galaxies with multiple spin reversals ({\it Group c}: 8/25),  and galaxies with regular rotation ({\it Group e}: 6/25). Noteworthy is that 44\% of {\it Group b} and all galaxies of {\it Group c} have at least one component in long-axis rotation (52\% of the full sample). Recent simulation studies predict that the emergence of prolate morphology and long-axis rotation is linked to the formation of massive galaxies \citep{2017ApJ...850..144E, 2018MNRAS.473.1489L, 2022MNRAS.509.4372L}. M3G data contribute by showing that the long-axis rotation is prevalent among massive galaxies, in the sense that it exists even if it is not a dominant component.

The velocity maps of the M3G {\it Group c} galaxies are remarkable for atypically large number of kinematic features. Galaxies up to $5-6\times10^{11}$ $M_\odot$ generally exhibit at most two different stellar kinematic components, with respective spins aligned with either the principal major or minor photometric axis \citep[e.g., NGC\,4365, the quintessential example of a KDC galaxy;][]{2001ApJ...548L..33D}. While KDC galaxies are also present in the M3G sample of galaxies, we reveal here systems that have much more intricate velocity maps, as well as multiple spin reversals. The most remarkable cases are 

\begin{itemize}
\item PGC\,019085 (MS-3L), with a central KDC, outer counter-rotation and an additional component with long-axis rotation, 
\item PGC\,046832 (MS-5L), presented earlier (Fig.~\ref{f:ms}) with five spin reversals, 
\item PGC\,047590 (MS-2L), with prograde and retrograde {\it long-axis} rotations, and
\item PGC\,047752 (MS-4L), with a central $5\arcsec$ region showing both short and long-axis rotations, and a retrograde short-axis rotation at larger radii, which twists strongly, but incompletely, towards long-axis rotation at the edge of the MUSE field.
\end{itemize}

\begin{figure}
    \includegraphics[width=\columnwidth]{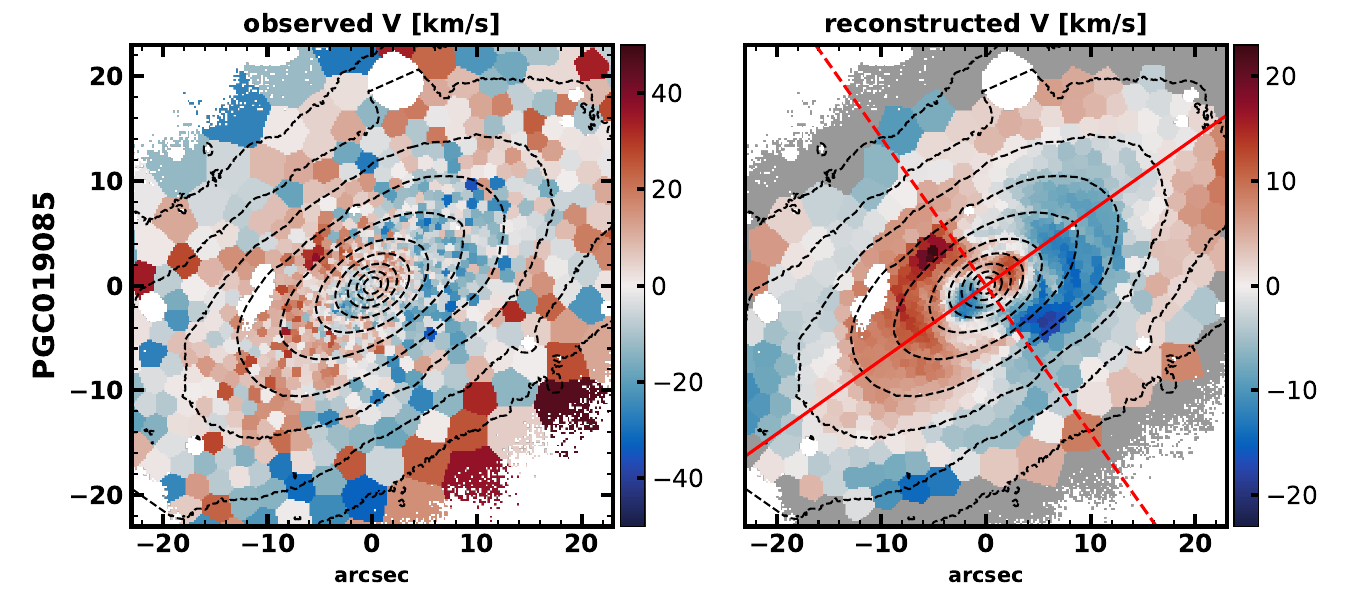} 
    \includegraphics[width=\columnwidth]{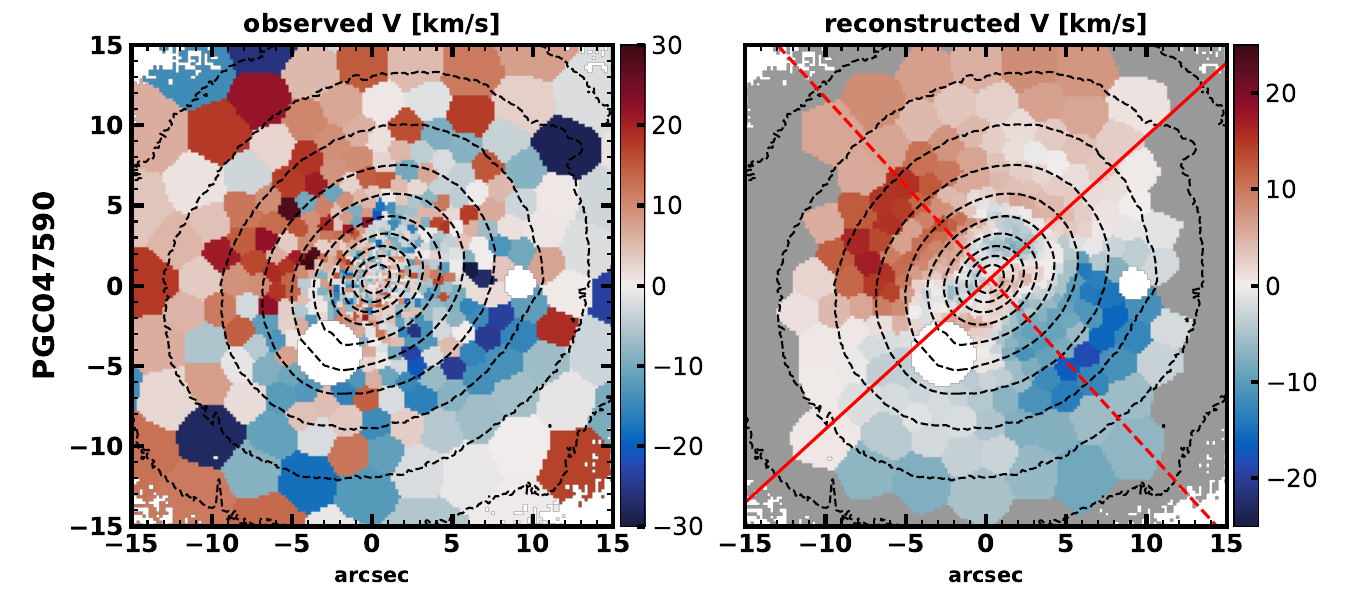} 
    \includegraphics[width=\columnwidth]{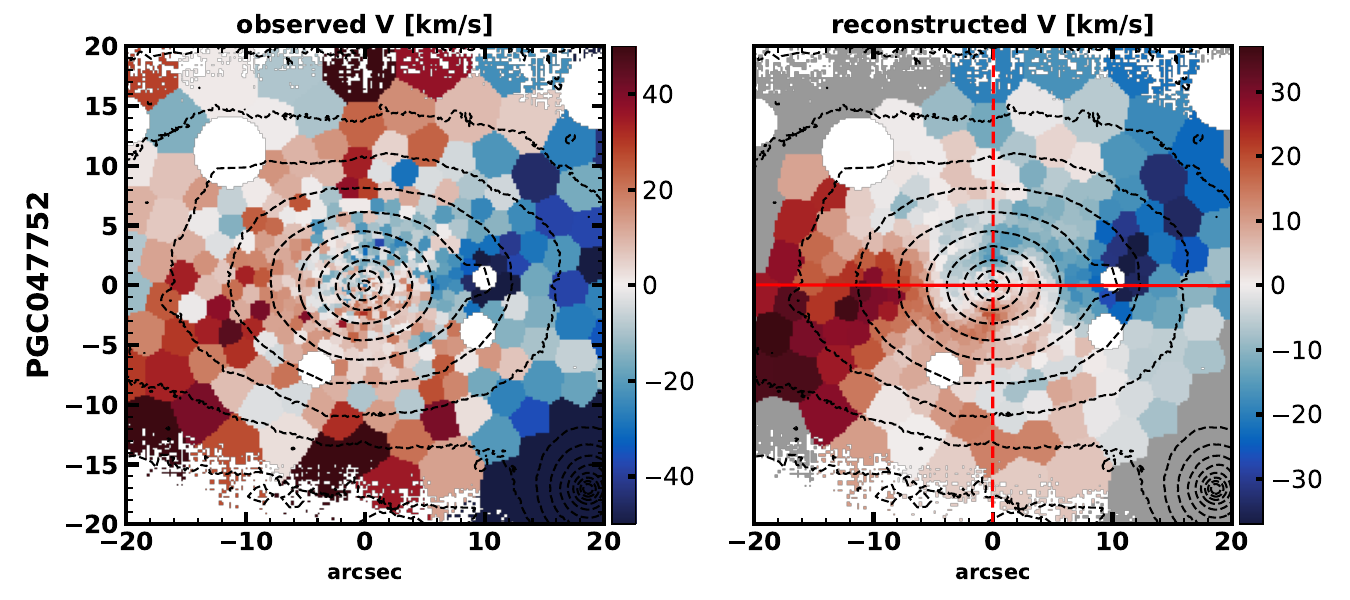} 
      \caption{M3G galaxies with extreme velocity maps, where the observed map is on the left and the right panel shows the reconstructed velocity maps using low-order \textsc{kinemetry} coefficients, highlighting only the rotational component. Note that the colour bars have somewhat different scales, tuned to highlight the features on individual maps. Red solid and dashed lines are the projected major and minor axes of the galaxy, respectively.  Kinematic features are best seen by following the spin reversals along the major and minor axes. Another extreme, PGC\,046832, was already shown in Fig.~\ref{f:ms}, which has five visible changes of the rotational orientation.}
         \label{f:ms_exp} 
   \end{figure}

\noindent We shows these systems (except PGC\,046832) in Fig.~\ref{f:ms_exp}, illustrated by their observed velocity maps and \textsc{kinemetry}-reconstructed velocity maps (see also Figs.~\ref{f:kinem} and S3 for more details). Another potential case is PGC\,043900 (see \textsc{kinemetry}-reconstructed velocity map in Fig.~S3). It is classified as a non-rotating {\it Group a} galaxy, but despite the low speeds, using the harmonic analysis and averaging the velocities along the ellipses, one can see hints of both prograde and retrograde short-axis rotations throughout the main body of the galaxy, as well as evidence for long-axis rotation confined to the central 10\arcsec.

Complex velocity maps of equilibrium models are possible only in non-axisymmetric objects, as the observed velocity map is a luminosity-weighted average of the stellar orbits that build the galaxy \citep{1987ApJ...321..113S}. In addition, the shape of the velocity map will also depend on the projection, or the viewing angle, with respect to the orientation of the galaxy \citep{1988FranxThesis, 1991AJ....102..882S, 1994MNRAS.271..924A}. In his analysis of triaxial models based on the ``perfect ellipsoid'' -- a St\"ackel potential form \citep{1985MNRAS.216..273D} -- \citet{1991AJ....102..882S} demonstrated that strong kinematic twists, long-axis rotation, and KDCs arise naturally from the presence of three distinct families of tube orbits and the effects of varying viewing angles. This is true even when rotation is restricted to {\it only} prograde streaming along the short- and long-axis tubes. Due to projections effects, the \citet[][Fig.~3]{1991AJ....102..882S} velocity maps included classic KDCs of the NGC\,4365 type, as well as predicted the possibility of observing types of long-axis counter rotation present in PGC\,047590 (Fig~\ref{f:ms_exp}). The velocity maps of real galaxies are, however, more complex than the ``perfect ellipsoid'' models with streaming around two axes, because we observe both prograde and retrograde rotations. 

Evidence for stellar counter-rotation in oblate disk galaxies is abundant \citep[starting from the archetypal NGC\,4550;][]{1992ApJ...394L...9R}. They can be traced to the accretion of gas on retrograde orbits \citep[e.g.][]{1992ApJ...401L..79B, 1994ApJ...432..575M, 2004A&A...424..447P, 2013A&A...549A...3C}, mergers \citep{2011MNRAS.416.1654B}, or potentially to cold streams \citep{2014MNRAS.437.3596A}. In massive ETGs, which rotate more slowly, it is much more difficult to directly extract the kinematics of different components by spectroscopically disentangling them. However, one can investigate the internal structure of ETGs using dynamical models, especially with orbit-based modelling \citep{1979ApJ...232..236S}. A conceptual proof of this approach was presented with a dynamical model of NGC\,4550 \citep{2006MNRAS.366.1126C}, demonstrating that the observed kinematics can be reproduced by placing $\sim50$\% of the mass into the counter-rotating component. Another case was the classic KDC in NGC\,4365, which features two orthogonal spin axis coinciding with the major and minor photometric axes. Using triaxial dynamical models, \citet{2008MNRAS.385..647V} revealed that to reproduce the kinematics, the internal structure requires three different orbital components with streaming: prograde and retrograde short axis tubes and (prograde) long axis tubes\footnote{The model also requires box orbits, but as they do not carry a net angular momentum they do not contribute to the velocity map. Instead, they are needed to reproduce the even moments of the LOSVD and the mass distribution.}. Further studies using Schwarzschild's method showed that galaxies with KDCs or counter-rotating disks can be similarly reproduced by superposition of orbits \citep[e.g. IC1459, NGC\,4473, NGC\,5813 by][]{2002ApJ...578..787C, 2007MNRAS.379..418C, 2015MNRAS.452....2K, 2021MNRAS.504.5087M}. These studies showed that even what is observed as non-rotation can be explained as a (luminosity or mass weighted) combination of prograde and retrograde orbits. 

One of the most complex-looking M3G galaxies, PGC\,046832 (Fig.~\ref{f:ms}), was modelled in a similar way with triaxial Schwarzschild models by \citet{2021MNRAS.508.4786D}, who concluded this galaxy is built of prograde and retrograde short axis tubes (42\% of the mass), prograde and retrograde long axis tubes (34\%) and box orbits (23\%). Crucially, all four families of streaming orbits are needed to reconstruct the five spin reversals observed in the velocity map. Therefore, it is reasonable to assume that all other multi-spin galaxies will require a similar orbital set-up to reproduce their velocity maps, in addition to the projection effects due to their triaxial nature and specific viewing angles. A working hypothesis is that the orbital families permeate through each other and the full body of the galaxy, while the features in velocity maps depend on the relative spatial contributions of various orbital families, their prograde and retrograde fractions, the viewing angles and projection effects \citep{1991AJ....102..882S,1994MNRAS.271..924A}. It follows that features visible in the velocity maps (e.g. a KDC) are not a specific well defined entities in real space, in the sense of being a physical and spatially confined component, separated from the rest of the galaxy body \citep{2008MNRAS.385..647V, 2015MNRAS.452....2K, 2021MNRAS.504.5087M}. 

The simplest explanation of retrograde motions (counter-rotation) is that it originates from accretion of external material. In case of oblate counter-rotating disks, the accretion of gas is sufficient \citep{2014MNRAS.437.3596A}. For triaxial and massive galaxies, strong tidal interactions and mergers are the most likely explanation for the observed richness of kinematics \citep{2010ApJ...723..818H, 2021A&A...647A.103E}, especially when they are gas-free mergers of similar mass galaxies \citep{2014MNRAS.445.1065R, 2014MNRAS.444.3357N, 2019MNRAS.489.2702F}. This is supported by the variety and apparent randomness of the observed velocity maps: velocity maps of massive galaxies rarely exhibit significant similarities, even when the galaxies have the same number of spin reversals. Furthermore, multi-spin velocity maps, and especially those with more than two spin reversals, occur in brighter (more massive) galaxies, which are more likely to experience gas-free mergers. Finally, the fine-tuned balance among orbital families that might even lead to the cancellation of detectable streaming motions requires several merging events. Such systems are the rarest (3\% in ATLAS$^{\rm3D}$ survey and $<1\%$ in M3G), and happen in the most massive galaxies, which again are more likely to experience multiple minor and major mergers \citep{2014MNRAS.444.3357N}. 

Conversely, velocity maps of RR galaxies resemble each other to a much higher degree, which makes sense if their formation is predominantly through dissipative processes, which include star formation or and disk formation. Their subsequent passive evolution might transform them into massive fast rotator ETGs, but not into multi-spin slow rotators. Therefore, multi-spin structures on velocity maps are possible only if galaxies are triaxial, allow for streaming along the long and the short axis, and are formed through interactions and mergers.

\subsection{Slow rotation at the highest masses}
\label{ss:sr}

All M3G BCG are slow rotators. Additionally, all non-BCG M3G members of Abell\,3562 cluster are slow rotators, as well as the two brightest non-BCGs in Abell\,3558. Abell\,3562 is somewhat special as its four brightest members have similar brightness \citep[e.g. there is no typical brightness gap between the BCGs and non-BCGs,][]{2014ApJ...797...82L}, the cluster does not seem settled and several galaxies are bright X-ray sources (see Section~\ref{ss:sample}). Still, the M3G sample follows a trend: slow rotators are BCGs, possibly include one or two highest ranked galaxies in a cluster, and they are found in regions more densely populated with other galaxies (in comparison with fast rotators). This is consistent with previous studies \citep{2013ApJ...778L...2C, 2013ApJ...778..171J, 2015MNRAS.454.2050F, 2017AJ....153...89O, 2017ApJ...844...59B, 2019arXiv191005139G}, where BCGs of massive clusters are exclusively slow rotators, and only a few other galaxies within the cluster, typically at centres of sub-clusters groups, can also be slow rotators \citep{2013MNRAS.429.1258D,  2014MNRAS.443..485F, 2014MNRAS.441..274S, 2019arXiv191106103G}. 

In this respect, our results support the existence of the kinematic morphology - density relation \citep{2011MNRAS.416.1680C,2013ApJ...778L...2C}. While the environment might not be the main driver \citep{2017ApJ...844...59B,2017ApJ...851L..33G, 2021MNRAS.508.2307V}, the mass function of slow rotators depends on the environment \citep[][]{2019arXiv191005139G}. The M3G sample contains some of the largest and brightest galaxies and extends the current sample of massive galaxies to $K_s\approx-27$ absolute magnitude. As a consequence we report that galaxies brighter than $-26$ mag are almost exclusively slow rotators (Fig.~\ref{f:lr_mag}). This level of brightness is only achieved by BCGs, which also do not belong to the general population (or the luminosity function) of massive galaxies \citep[e.g.][]{2010ApJ...715.1486L}. The M3G sample illustrates the interconnectedness of galaxy mass growth and the loss of angular momentum. Slow rotation of BCGs is naturally explained through repeated gas-poor major mergers, as well as the frequent accretion of smaller satellites at their ``privileged'' position at the bottom of the cluster potential well. Such mergers contribute to the reduction of angular momentum, growth of galaxy mass and diversification of the types of orbits, and explain the variety of kinematic features on velocity maps.

\subsection{Fast rotation at the highest masses}
\label{ss:fr}

Using \textsc{kinemetry} we established that there are six galaxies which are consistent with regular rotation. All these galaxies can unambiguously be classified as fast rotators. Their existence is not entirely surprising: among the nearby ETGs, regular rotation is found in galaxies as massive as $5-6\times10^{11}$ $M_\odot$ \citep{2020A&A...635A.129K} and the MASSIVE survey reports fast rotators up to $10^{12}$ $M_\odot$ \citep{2017MNRAS.471.1428V}. None of the M3G fast rotators is a BCG. Instead, they are by brightness all lower ranked galaxies in their clusters. The only exception is the second ranked galaxy in Abell\,3556, PGC\,046785, which, as argued before, is marginally a fast rotator within its effective radius. All other M3G fast rotators have regular velocity maps, suggesting the existence of a single dominant orbital family of short-axis tubes. 

This assessment is based on $\xi_3$ values of M3G fast rotators, which are all, except one, strongly negative, implying an anti-correlation $V/\sigma - h_3$ (Fig.~\ref{f:vs_h3}, left). Simulations predict that such galaxies have high fraction of short-axis tubes, dominating the orbital distribution, and are likely to have experienced gas rich mergers producing fast rotators \citep{2009ApJ...705..920H, 2014MNRAS.444.3357N, 2014MNRAS.445.1065R, 2019MNRAS.489.2702F}.  An exception is PGC\,047177, which therefore must have more complex orbital structure, not visible in the velocity map.

Fast rotators are galaxies less bright than $-26$ mag ($K_s$-band), and approximately less massive than $8\times10^{11}$$M_\odot$. As there is a clear relation between the formation of slow rotators and mergers of certain type \citep{2010ApJ...723..818H, 2011MNRAS.416.1654B,2014MNRAS.444.3357N}, this brightness (mass) limit possibly defines an approximate threshold above which formation of fast rotators is less likely.

\subsection{BCGs and their massive companions}
\label{ss:mcc}

There is ample evidence that BCGs are different from other galaxies \citep[e.g.][]{1977ApJ...212..311T, 1996ApJ...465..534G, 2014ApJ...797...82L, 2018MNRAS.477..335L, 2020MNRAS.496.1857L}, but there are also similarities between BCGs and the highest mass non-BCGs. Kinematically, for example, short-axis rotation is found both in BCGs and non-BCGs, and there are also BCGs that have very small kinematic misalignment angles \citep{2018MNRAS.477.5327K,2018MNRAS.479.2810E}. We showed that there are non-BCGs that are slow rotators and have multiple-spin reversals, but they are rare. Regular rotation and fast rotators are only found in non-BCGs, and long-axis rotation predominantly occurs in BCGs: 10/14 (71\%) compared to 3/11 (27\%) in non-BCGs. 

We also presented another difference between BCGs and non-BCGs: a radially varying $k_0$ term. As discussed in Section~\ref{sss:var_sys} and App.~\ref{app:vsys}, in most cases this is likely an indication of the existence of another kinematic component at outer radii. We did not present the evidence here, which we postpone for a future paper, but galaxies that show large $\delta k_0$ variation also have radially raising velocity dispersion profiles (see e.g. velocity dispersion maps of PGC\,015524 and PGC\,047202 on Fig.~\ref{f:kinem}). Rising velocity dispersion profiles are found in massive galaxies \citep{2018MNRAS.473.5446V} and are specific for BCGs and cD galaxies \citep{2007MNRAS.378.1507B, 2013ApJ...765...24N} where velocity dispersion reaches the value of the velocity dispersion of the cluster galaxies. An example is NGC\,6616 showing radial variation in the mean velocity and the velocity dispersion, both of which approach the mean cluster values \citep{2015ApJ...807...56B}. The authors explain such high velocity dispersion values by assuming that outer stellar component of the BCGs is build up from stars accreted in minor mergers or stripped from cluster galaxies. Our data do not extend to large enough radii, but we see a similar trend where the velocities, parameterised by $V_{\rm sys} + k_0$, rise towards the mean cluster velocity values. Therefore, it is possible that the $\delta k_0$ variations trace an increasing contribution of a component made up of externally accreted stars. As this component likely originates from tidally stripped material of other cluster galaxies, it is (nearly) at rest with respect to the cluster mean velocity and velocity dispersion based on the motions of all cluster galaxies. In this sense, the kinematics of this components could be different from the main body of the BCG, which has its specific location and motion within the cluster. 

Numerical simulations \citep{2024ApJ...968...96H} find the increase in the velocity dispersion at outer radii to be related to the high fraction of stars formed {\it ex situ} and higher velocity anisotropy which these stars carry. The same authors argue that the variety of the outer velocity dispersion profiles in galaxies reflects their diverse assembly histories. We will investigate the velocity dispersion radial profiles and their connection to $\delta k_0$ in an upcoming study. Nevertheless, we conclude that if one can establish a connection between the radial $k_0$ increase and accretion of stars born elsewhere, which form an additional kinematic component, one can add another evidence to different formation paths of BCGs and non-BCGs.

\section{Conclusions}
\label{s:con}

The observed stellar kinematics are a reflection of the richness of the underlying orbital distributions which build up galaxies. This is seen in the orderly velocity maps of regular rotators, which comprise the majority of galaxies, and in the kinematic complexity observed in the velocity maps of the most massive systems. This interpretation relies on the hypothesis that galaxies are built of several orbital families, which leave their imprint on the velocity maps, such as the short-axis and long-axis tubes, rotating around the minor- and the major-axis of the galaxy, respectively. There are also orbits that have no angular momentum, and although they do not leave imprint on the velocity maps they participate in shaping the even kinematical moments, such as galaxy morphology and the velocity dispersion. 

In this work we described the observational setup and data associated with the M3G Survey, a MUSE GTO program, targeting several of the brightest (most massive) galaxies within $\sim 200$ Mpc. From observations of 25 such galaxies, divided into 14 BCGs and 11 non-BCGs, we extracted stellar kinematics and analysed their velocity and $h_3$ maps. We demonstrate that velocity maps of massive galaxies can show several reversals in the orientation of the visible rotation, many more compared to what was known for less massive galaxies ($<5\times10^{11}$ $M_\odot$), including classical KDCs. In the extreme case of PGC\,046832, there are five different spin-flips in projection. Furthermore, the majority of BCGs galaxies and several luminous non-BCGs have at least one component that shows long-axis rotation (rotation around the major photometric axis). These complex velocity maps reinforce the notion that massive galaxies have non-axisymmetric shapes and contain several orbital families, but stars need to stream in both prograde and retrograde directions within their parent orbital families. 

We find that there are massive galaxies that have very simple velocity maps (regular rotation) and that they are all fast rotators. They are non-BCGs, and also low ranked (in luminosity) within their clusters, in spite of all sample galaxies being among the brightest in the nearby universe. Fast rotating galaxies also show a strong anti-correlation between $V/\sigma$ and $h_3$ parameters, which, in comparison with numerical simulations, suggests their orbital distribution is simple and dominated by short-axis tubes only. 

Some BCGs show kinematic evidence for existence of an additional component not fully synchronised with the main component. The difference in the kinematics points to an external origin of this component, possibly being made of stars stripped from other cluster galaxies. This is only a tentative conclusion which will be explored more in a future paper discussing the velocity dispersion and the $h_4$ moment maps of M3G galaxies. 

Overall, our findings support the general paradigm of formation of massive galaxies \citep[e.g.][]{2010ApJ...725.2312O, 2016ARA&A..54..597C, 2017ARA&A..55...59N}, providing further constraints that the formation of the most massive galaxies is governed by merging, typically gas-poor, but having diverse characteristics such as the mass ratios, orientation of the orbital angular momentum and frequency depending on the local environment. Velocity maps remain a very good indicator of the complexity of evolutionary histories, especially at the tip of the galaxy mass distribution.

\section*{Data Availability}

The Supplementary material is accessible at \url{https://doi.org/10.5281/zenodo.18459530}

\begin{acknowledgements}
We thank Ilaria Pagotto and Mark van den Brok for contributions to data analysis and discussions at early stages of the project. DK acknowledges support from the grant: KR 4548/4-1 of the Deutsche Forschungsgemeinschaft. L.A.B. acknowledges support from the Dutch Research Council (NWO) under grant VI.Veni.242.055 (\url{https://doi.org/10.61686/LAJVP77714}). We have used the services from SIMBAD, NED, and ADS as well as the Python packages NumPy \citep{2007CSE.....9c..10O}, Matplotlib \citep{2007CSE.....9...90H}, SciPy \citep{2020NatMe..17..261V}, and Astropy \citep{2018AJ....156..123A}, for which we are thankful. The MUSE observations were taken within the following observing programmes: 094.B-0592, 095.B-0127, 096.B-0062, 097.B-0776, 098.B-0240,099.B-0148,099.B-0242, 0102.A-0327.

\end{acknowledgements}

%
\bibliographystyle{aa} 

%

\begin{appendix}

\section{Extraction of satellites and their kinematics}
\label{app:sat}

Figure S2 in the Supplementary Material presents three examples of M3G galaxies with fore- or back-ground satellites. Plots for each galaxy are distributed over two columns, where the first row shows the map of the $\chi^2$ used to allocate whether a bin belongs to the main galaxy, satellite galaxy or both, and the distribution of these bins (see Section~\ref{ss:bin} for details).  PGC\,047355 is the only case in the sample where both the galaxy and the satellite kinematic components can be disentangled at all bins. In all other cases, the central regions of the satellites are completely dominating its light and we are not able to obtain any information about the main galaxy kinematics. The spectra in which we can separate the components are typically confided to a small region at the boarders of the satellites. An exception is PGC\,099188*, used as an example in the main text (Section~\ref{ss:bin}), where the multi-component pPXF fit is able to separate both the satellite and the main galaxy component over the most of the field, except in the central $10\arcsec$ of the satellite.  Another extreme is shown by PGC\,065588 and its satellite in the lower left corner, where only a handful of bins at the edges of the satellite can be used to extract both kinematic components, while for PGC\,099522 the two satellites completely dominate one side of the galaxy and the main galaxy kinematics can not be extracted. Nevertheless, in most cases, the two component fitting procedure yielded a much increased region with kinematics of the main galaxy. Furthermore, it provided a natural and an optimal way of finding bins at the edges of the satellites that still contain significant fraction of the main galaxy light to extract its kinematics.  

There are 13 M3G galaxies with satellites whose kinematics we attempt to separate from the main galaxy. For six M3G galaxies, there is only one satellite in the MUSE FoV. Five M3G galaxies have two interlopers, while PGC\,003342 galaxy has three satellites, and PGC\,047202 has seven. PGC\,047202 is the largest galaxy in the sample and found in the centre of the densest cluster, but all of the seven satellites that we separate kinematically are found outside of the central 1 arcmin (central MUSE pointing)\footnote{Note that there are compact sources within the central arc minute, which could be nuclei of stripped galaxies.}. Therefore, the reason why we detect such a larger number of satellites compared to the rest of the sample is due to the fact that we mosaic this galaxy with a $2\times2$ MUSE pointings quadrupling the area compared to other galaxies. The location of satellites can be seen on Fig.~\ref{f:kinem}. In total, we extracted kinematics for 26 satellites, of which half was in the foreground and half in the background of the main galaxies (when averaged over all M3G sample galaxies). We do not show the kinematics of satellites \citep[see][for kinematics of PGC\,047202 satellites]{2023Msngr.191...11B}, but their kinematics is dominated by rotation, and they are all classified as fast rotators (Fig.~\ref{f:lr}). In Table S1 of the Supplementary Material we provide an overview of satellites and their location on the MUSE maps. We also provide their ellipticities and $\lambda_{Re}$ values shown in Fig.~\ref{f:lr}, obtained using the same method as for the main galaxies (Section~\ref{ss:am}).

\section{Correcting velocity maps for the effect of variable $k_0$ parameter}
\label{app:vsys}

   \begin{figure*}
   \centering
    \includegraphics[width=0.9\textwidth]{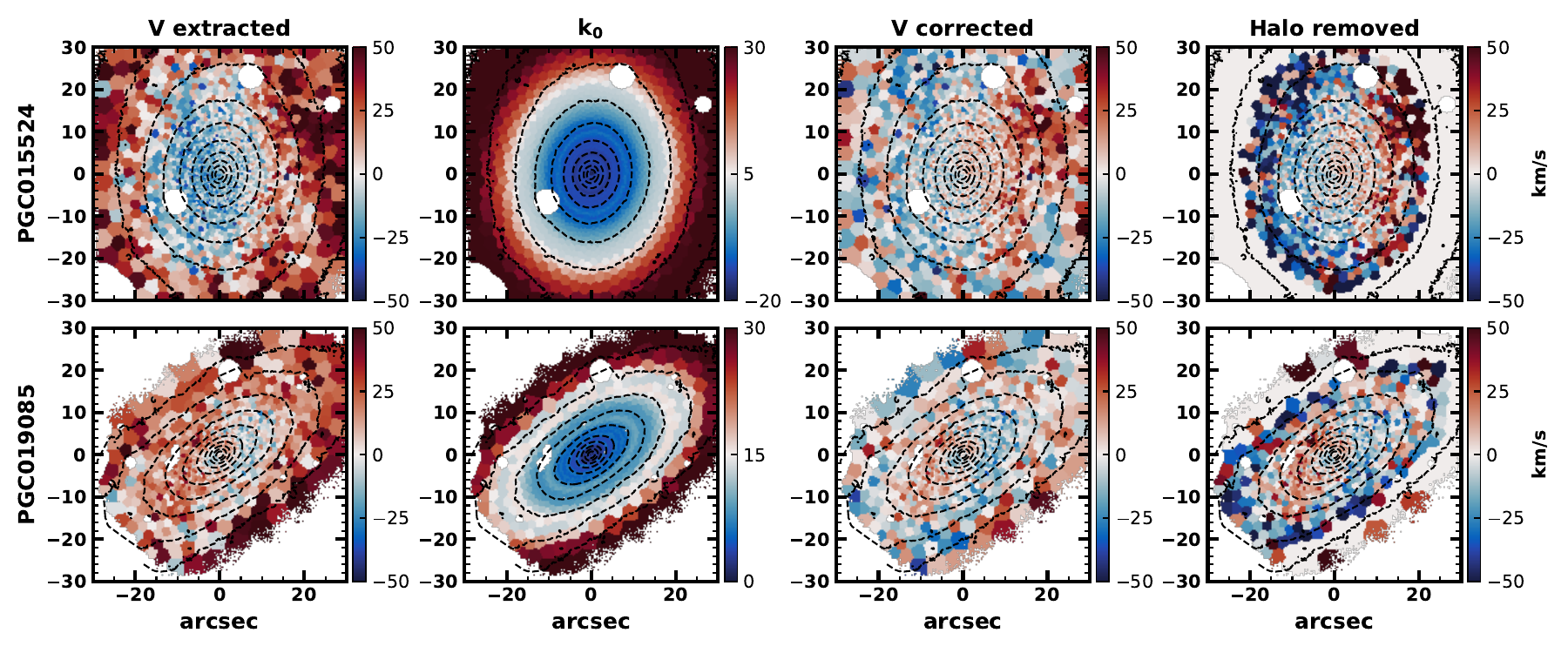} 
      \caption{The effect of removing the variable $k_0$ term is best visible in PGC\,015524 (top) and PGC\,019085 (bottom). {\it Left}: the original velocity map extracted using the pPXF from a binned MUSE cube. {\it Middle left}: maps of $k_0$ term in the harmonic analysis obtained with \textsc{kinemetry} on the original velocity maps. Note the different colour scales compared to other plots. {\it Middle right}: the corrected velocity map obtained by subtracting the $k_0$ value from each bin from the original velocity map. {\it Right:} velocity maps obtained through pPXF extraction as on the leftmost panels, but after first removing a representative halo spectrum (see text for details). Only bins with S/rN$>7$ are plotted. 
      }
         \label{f:vsys_rem}
   \end{figure*}

   \begin{figure}
    \includegraphics[width=\columnwidth]{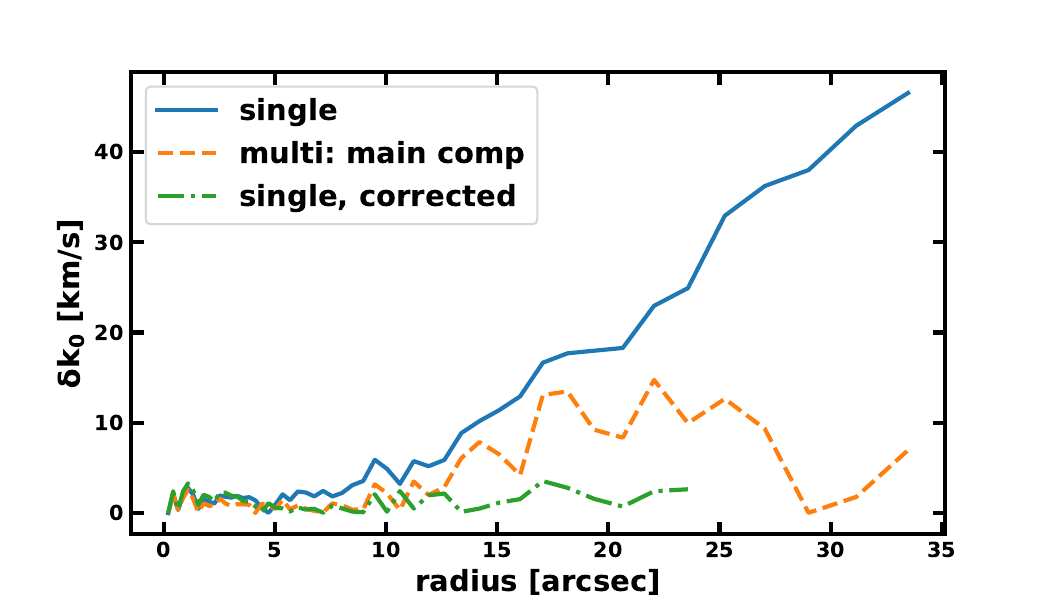} 
      \caption{Comparison of the absolute radial variation of $k_0$ from three kinematic extractions for PGC\,015524. The blue line shows $\delta k_0$ from the nominal extraction (same as in Fig.~\ref{f:vsys}). The orange dashed line shows $\delta k_0$ of the main kinematic components from a two components pPXF fit. The dashed dotted green line shows the $\delta k_0$ from the kinematic extraction after removing a representative spectrum of the secondary component from the cube. The variation in $\delta k_0$ seems to originate in the existence of a secondary component which dominates the outer parts. }
         \label{f:k0_kincomp}
   \end{figure}

The velocity measured by the pPXF is a combination of the peculiar velocity, $V_{\rm pec}$(x,y), specific for each bin and the velocity of the barycentre of the galaxy. The former is the velocity that describes the internal dynamics of the galaxy, and is directly related to the orbital distribution, while latter is a measure of the cosmological redshift of the galaxy \citep[see eq.~(3) and Section 2.2 in][]{2023MNRAS.526.3273C}. This is usually referred to as the systemic velocity, $V_{\rm sys}$. For clarity, we use an example of a two dimensional velocity map of a disk galaxy, described as: 
\begin{equation}\label{eq:d}
$$V_{\rm pec}(x,y) = V(R, \theta) = V_{\rm sys}(R) + V_{\rm rot}\cos(\theta)\sin(i),$$
\end{equation}
where $i$ is the inclination of the disk with respect to the plane of the sky (for $i=90\degr$ the galaxy is ``edge-on"), $\theta$ is the azimuthal angle measured from the major axis of length $R$ ($\theta=0\degr$), and $V_{\rm rot}$ is the amplitude of the rotation. The total amplitude of $V_{\rm rot}$ depends on the galaxy mass (i.e. traces circular velocity in a disk galaxy) and is modulated by the viewing angle, while its change across the field of the galaxy follows a cosine variation. If the galaxy redshift is precisely known, $V_{\rm sys}=0$, but this is rarely the case, and one measures $V_{\rm sys}\neq 0$ and a certain error $\Delta V_{\rm sys}$. The systemic velocity is not related to the internal kinematics or dynamics, and it should be constant across the galaxy. However, variations of measured $V_{\rm sys}$ for each $V_{\rm pec}$ are expected, and in practice one estimates a global value of $V_{\rm sys}$ as a median of the velocities across all independent spatial bins. 

The variation of $k_0$ across the field, as defined in \textsc{kinemetry} is, in principle, related to the uncertainties of the $V_{\rm sys}$, and its average value also provides an estimate for $V_{\rm sys}$. However, if the kinematics is not regular (a case different from the ideal disc of eq.~\ref{eq:d}), additional contribution might be present, tracing anther kinematic component.  This component could potentially not be fully bound to the main galaxy (e.g. following a recent merger or accretion), and retains a ``memory" of the cluster kinematics \citep{2015ApJ...807...56B}. In the outer regions where the main galaxy components becomes fainter, the contribution of the secondary components can bias the extracted kinematics, potentially detectable as a systematic shift in mean velocity. A natural way to detect this would be through a variation of $k_0$. 

Here we first present a way to remove the variation of $k_0$ across the field and determine the underlaying velocity of the main galaxy component, and subsequently we perform two different pPXF extractions to test the hypothesis that the variation of $k_0$ is due to existence of another kinematic component, detectable at large radii (beyond $R_e$ as suggested by Fig.~\ref{f:vsys}).

In order to determine the amplitude of $k_0$ variations, we extracted velocities along the fits to the isophotes, and not the best fitting ellipses (for a velocity map) as is usually done. In this way we trace the shape of the stellar distribution, and not the properties of the velocity map. We first run \textsc{kinemetry} on the MGE models of M3G galaxies (the same as used for masking in Sections~\ref{s:kin}). \textsc{Kinemetry} is run in its photometric mode (all harmonic terms) to determine the photometric position angle ($PA_p$) and flattening ($Q_p$). The MGE models are used as smooth representation of galaxies, with slowly varying $Q_p$ and $PA_p$, without satellite galaxies or compact objects.

The next step was to run \textsc{kinemetry} on the velocity maps, imposing the $PA_p$ and $Q_p$ in the extraction, and measuring the value of $k_0$, via harmonic analysis. We note that the ATLAS$^{\rm 3D}$  $k_0$ values, compared in Fig.~\ref{f:vsys}, were extracted in a slightly different way by running \textsc{kinemetry} on velocity maps only. The middle panels in Fig.~\ref{f:vsys_rem} show the maps of $k_0$, demonstrating a strong radial dependence, where $k_0$ rises towards the edge of the MUSE FoV, indicating a strong difference between the inner and outer regions of galaxies. The maps were subsequently subtracted from the original velocities producing the corrected velocity maps. Their examples are presented in the third column of Fig.~\ref{f:vsys_rem}, and show detailed velocity structures and clear improvements with respect to the original extractions. These maps are used in all subsequent analysis presented in the paper (Fig.~\ref{f:kinem}). 

As $k_0$ maps show the strongest difference at the outskirts of the MUSE field, we tested a hypothesis that this variation is caused by existence of another kinematic component, detectable at large radii. If this secondary component exists, by removing its representative spectrum we should be able to recover the kinematics of the primary component.  For both galaxies in Fig.~\ref{f:vsys_rem}, we selected spectra from $5-6$ regions, within $3\arcsec$ apertures, at the very edge of the MUSE field corresponding to high $k_0$ values (at least one in each quadrant of the field). We fit these outer spectra with pPXF and find their systemic velocities to be substantially different from that of the global galaxy spectrum (extracted within $1 R_e$). For PGC\,015524 the difference reaches 50 km/s and for PGC\,019085 about 35 km/s, comparable to the range of the respective $k_0$ values.  For each galaxy, we made a median stack spectrum based on the outer spectra, and subtracted it from each spectrum of the binned data cube. The pPXF fit on the whole data cube was executed as before, now extracting only two moments, $V$ and $\sigma$. In the last column of Fig.~\ref{f:vsys_rem}, we plot the resulting velocity maps for the two galaxies. The outermost bins are masked as the S/rN of the fit was low ($<7$). Nevertheless, both maps show clear velocity structures, essentially the same as those obtained by subtracting the $k_0$ parameter values. 

This test suggests that the origin of the large scales variation of $k_0$ is very likely due to another kinematic component which becomes dominant at large radii. We further tested this hypothesis by setting up pPXF to fit for two kinematics components, of different velocities and velocity dispersions, similar as in the case of fitting satellite galaxies (Section~\ref{ss:bin}). We performed this test on PGC\,015524 only, as the difference in kinematics between the components are small and large S/N is required for more robust results. After the pPXF extraction, we run \textsc{kinemetry} on the resulting velocity maps, estimated $k_0$ and $\delta k_0 = |k_0(0) - k_0(R)|$, and compared with the original $\delta k_0$ (e.g. in Fig.~\ref{f:vsys}). We also run \textsc{kinemetry} on the velocity map of the primary kinematic components from previous test (map on Fig.~\ref{f:vsys_rem}), and show all three $\delta k_0$ curves on Fig.~\ref{f:k0_kincomp}. 

The results of these tests are very suggestive that the variation in $k_0$ indeed originates in the existence of a secondary component which dominates the outer parts. The secondary components might not be in the equilibrium with the central (main galaxy) component (see discussion in Section~\ref{ss:mcc}), and could originate from accretion of satellites that distribute their stars at large radii and in the halo of the galaxy.

\section{Kinematic maps}
\label{app:kinmaps}

   \begin{figure*}
    \includegraphics[width=0.99\textwidth]{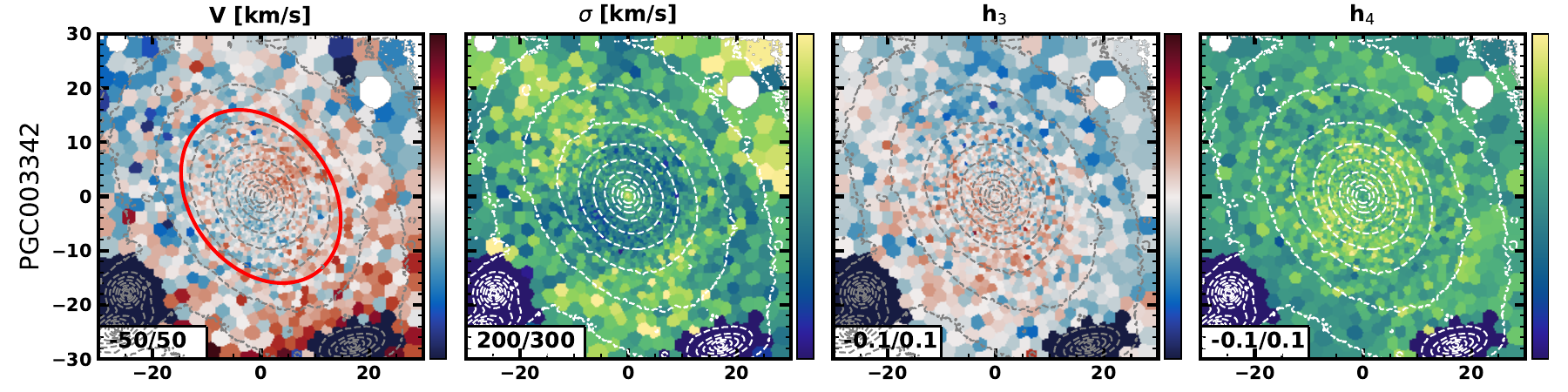} 
    \includegraphics[width=0.99\textwidth]{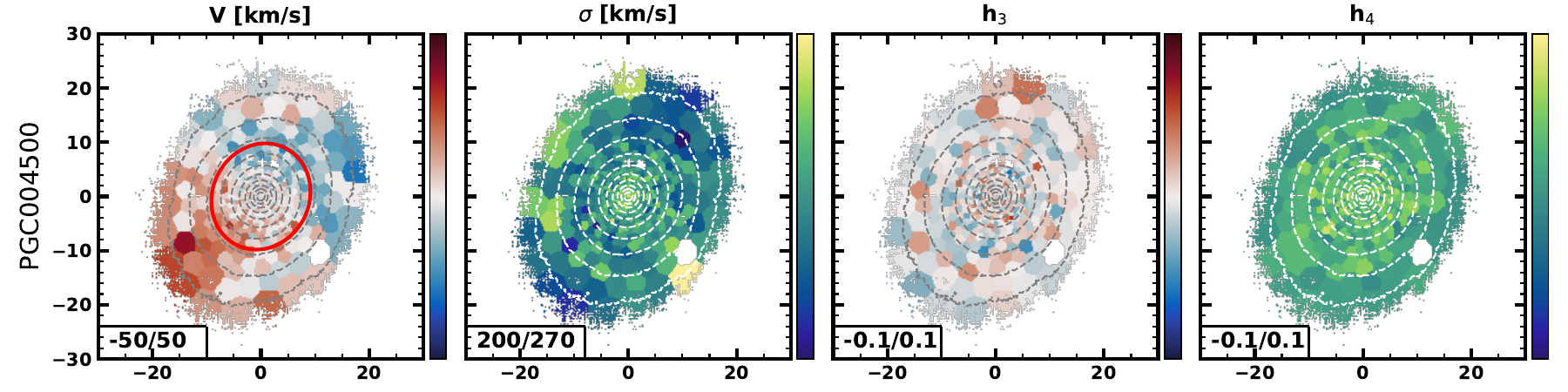} 
    \includegraphics[width=0.99\textwidth]{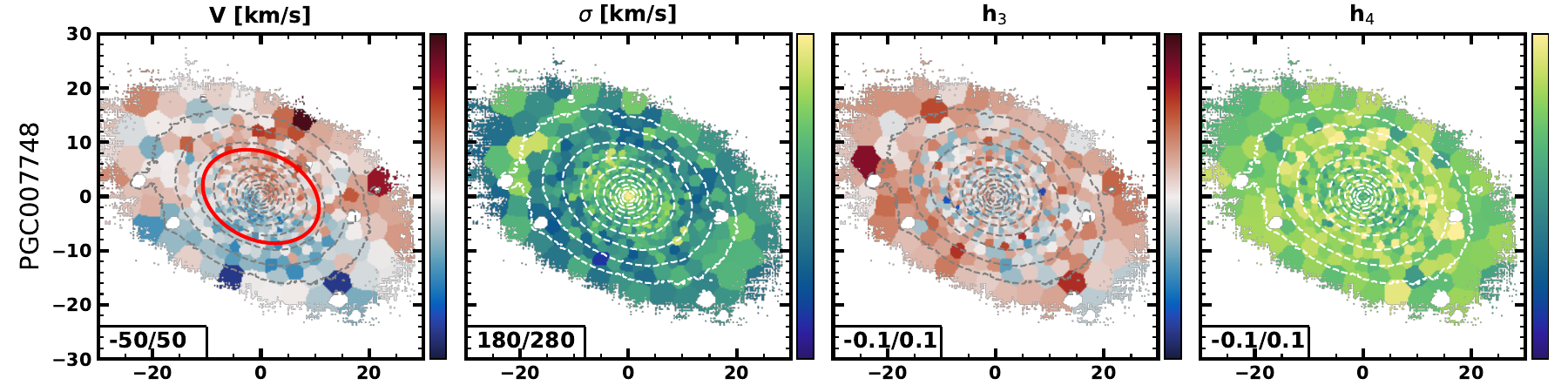} 
    \includegraphics[width=0.99\textwidth]{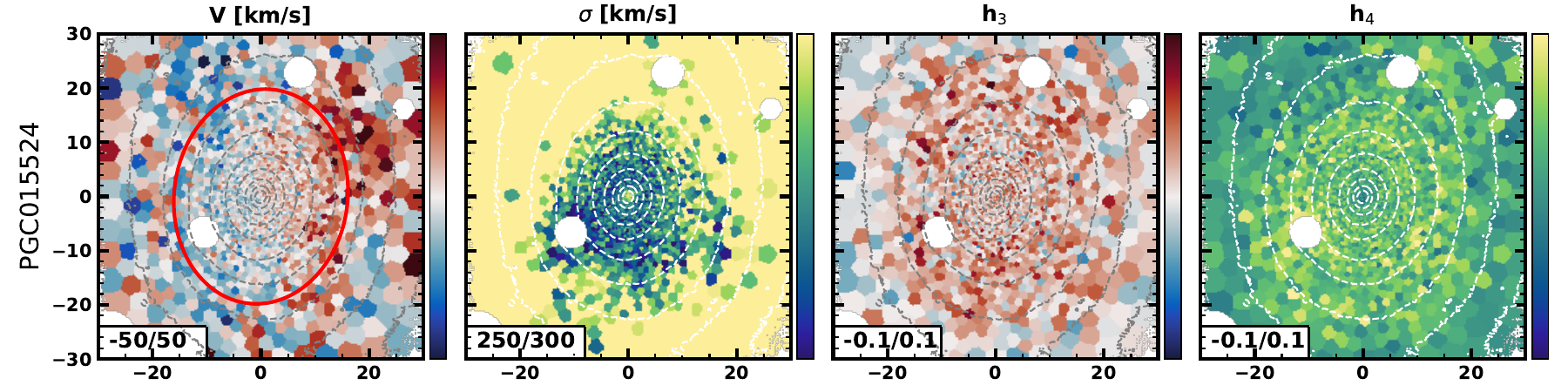} 
    \includegraphics[width=0.99\textwidth]{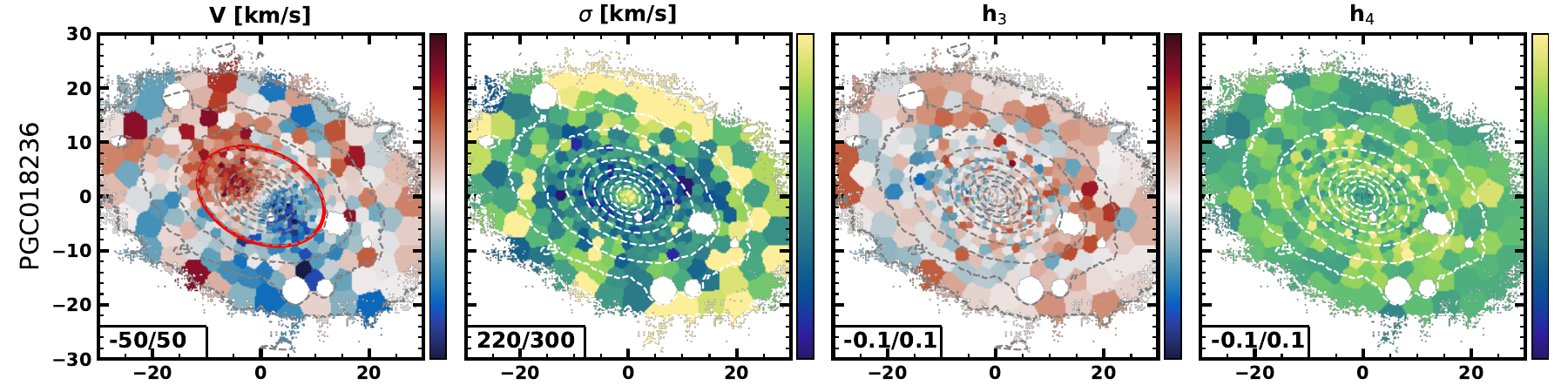} 
      \caption{Kinematic maps of M3G galaxies showing from left to right: the mean velocity, velocity dispersion, $h_3$ and $h_4$ moments. The values (in km/s for the mean velocity and velocity dispersion) in lower left corners correspond to maximum and minimum of the colour bars on the right of each plot. The red ellipse has the major axis length equal to $R_e$ and the median flattening of the galaxy within the effective radius (see Table~\ref{t:m3g}). It shows the elliptical aperture used to extract global {\it effective} properties (e.g. $\sigma_e$, $\lambda_{Re}$, average values of kinemetry parameters, etc).}
         \label{f:kinem}
  \end{figure*}

  \begin{figure*}
   \ContinuedFloat
    \includegraphics[width=\textwidth]{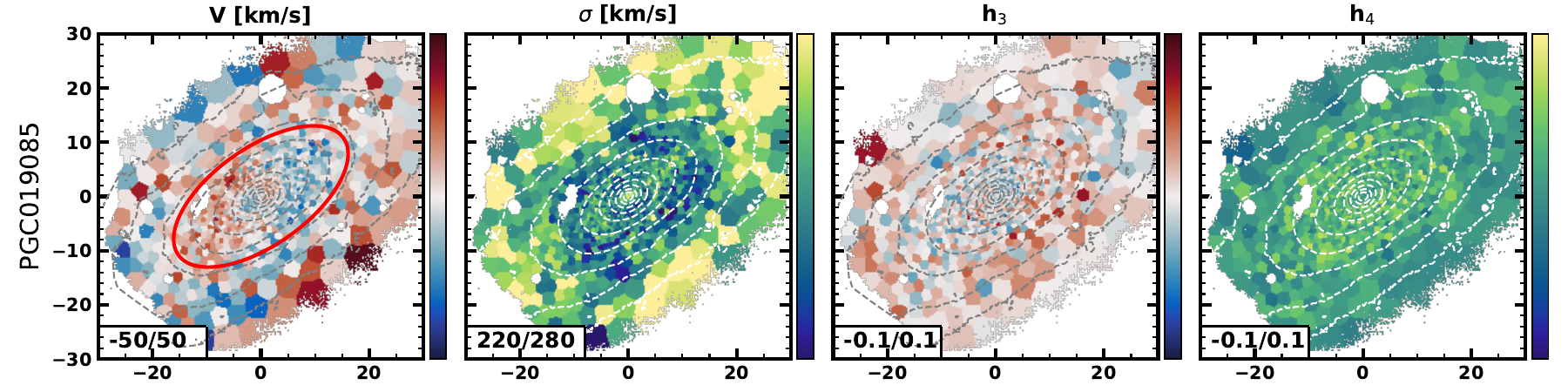} 
    \includegraphics[width=\textwidth]{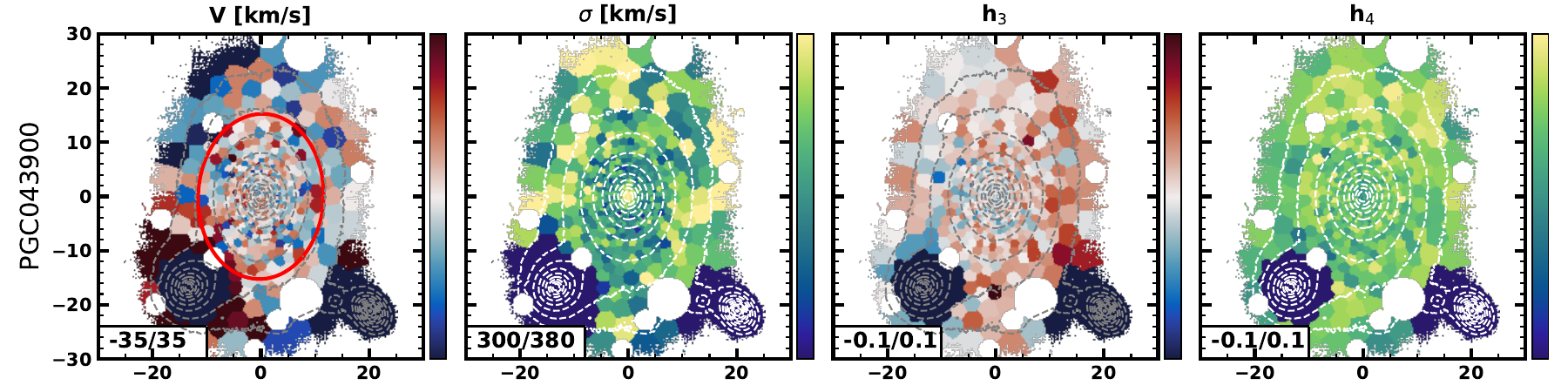} 
    \includegraphics[width=\textwidth]{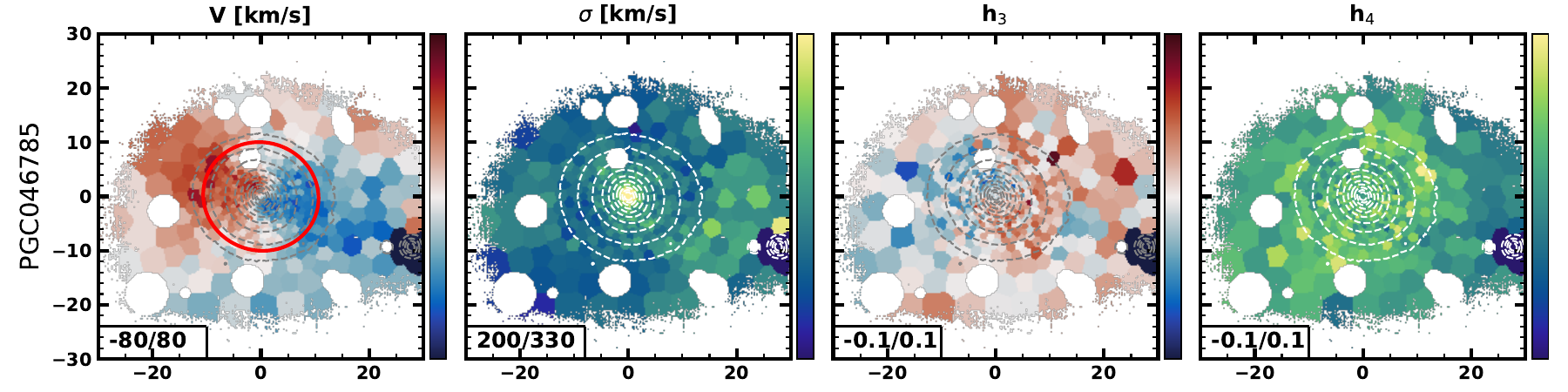} 
    \includegraphics[width=\textwidth]{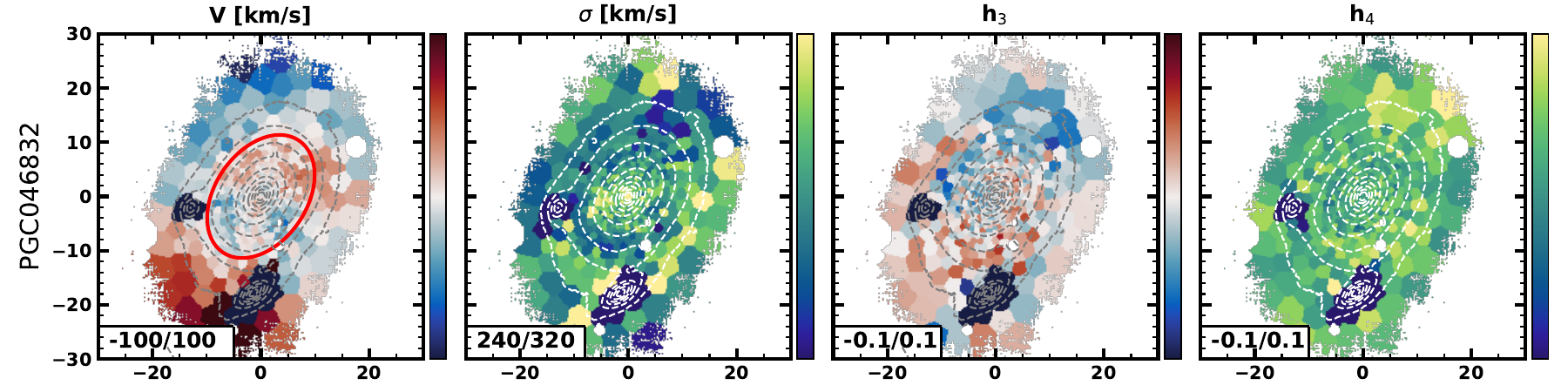} 
    \includegraphics[width=\textwidth]{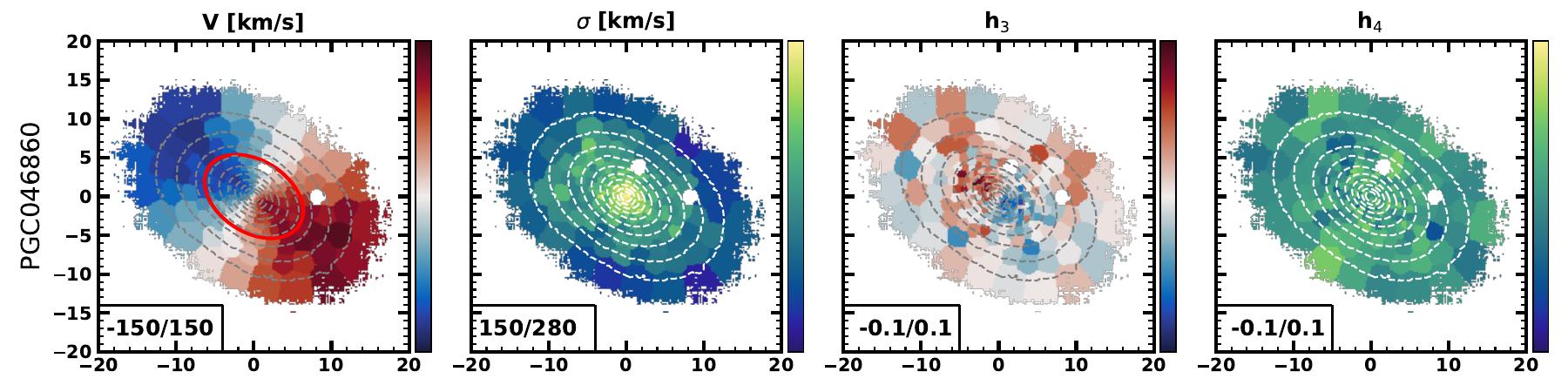} 
      \caption{Continued}
   \end{figure*}

  \begin{figure*}
   \ContinuedFloat
    \includegraphics[width=\textwidth]{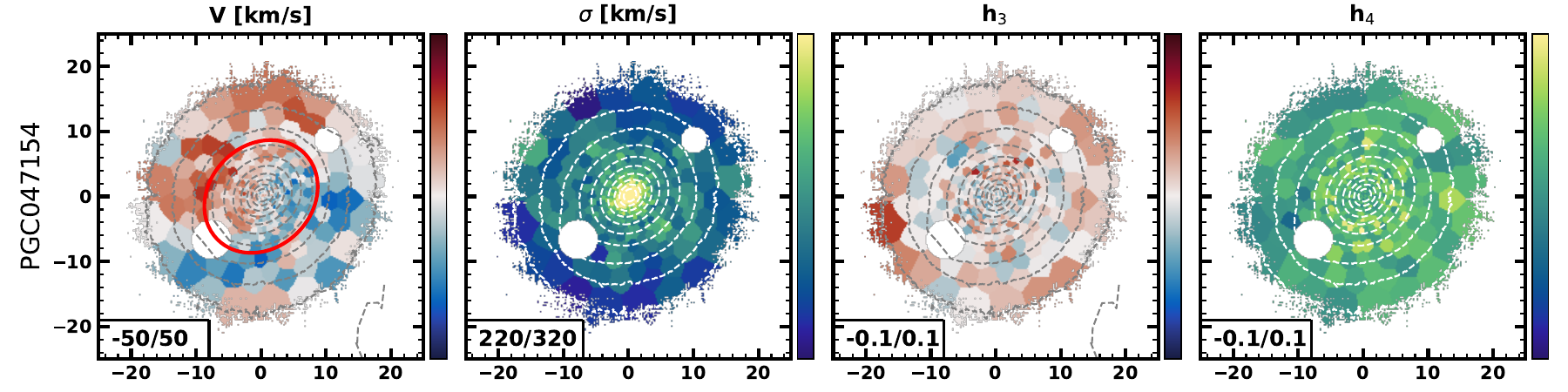} 
    \includegraphics[width=\textwidth]{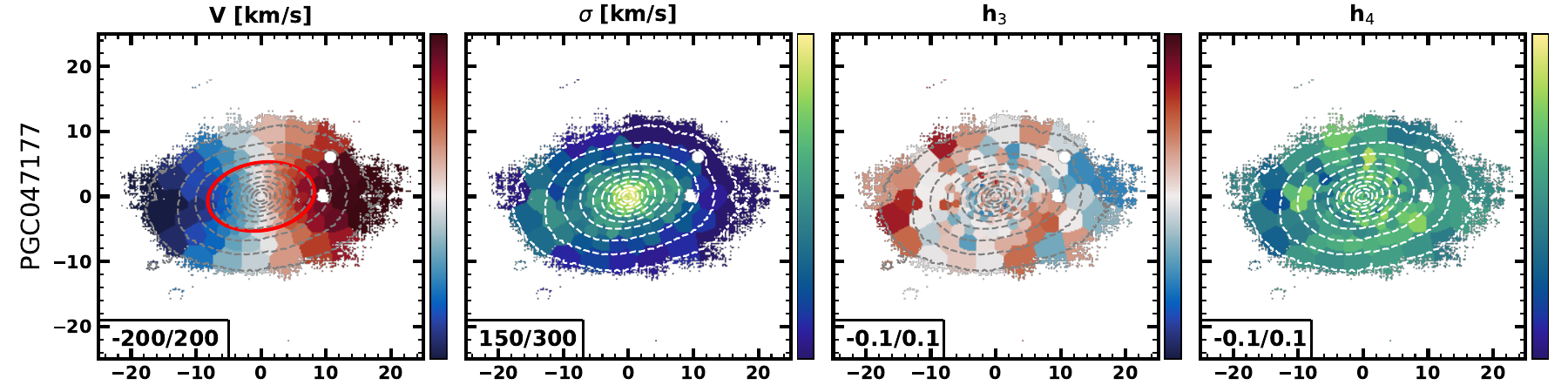} 
    \includegraphics[width=\textwidth]{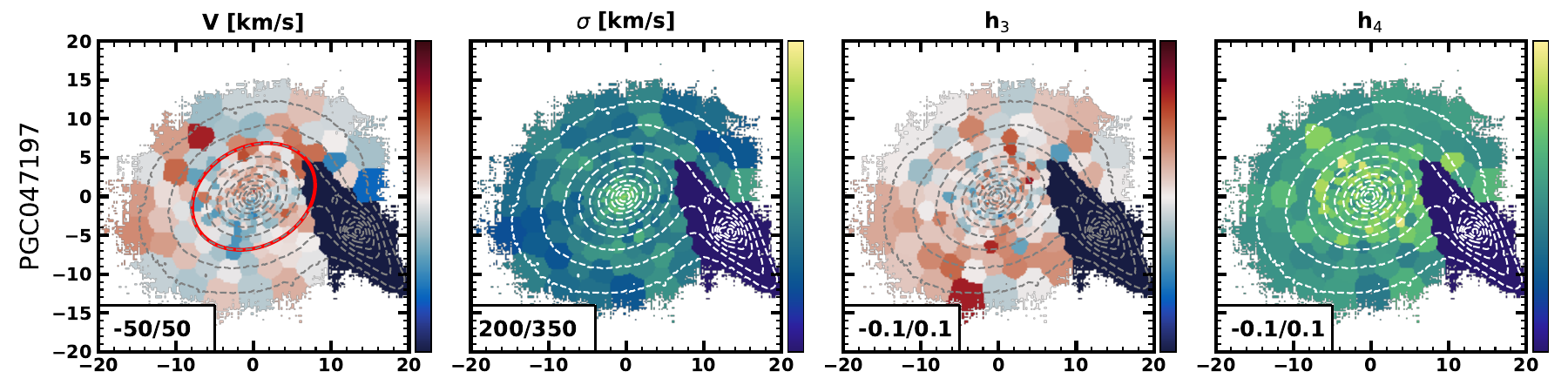} 
    \includegraphics[width=\textwidth]{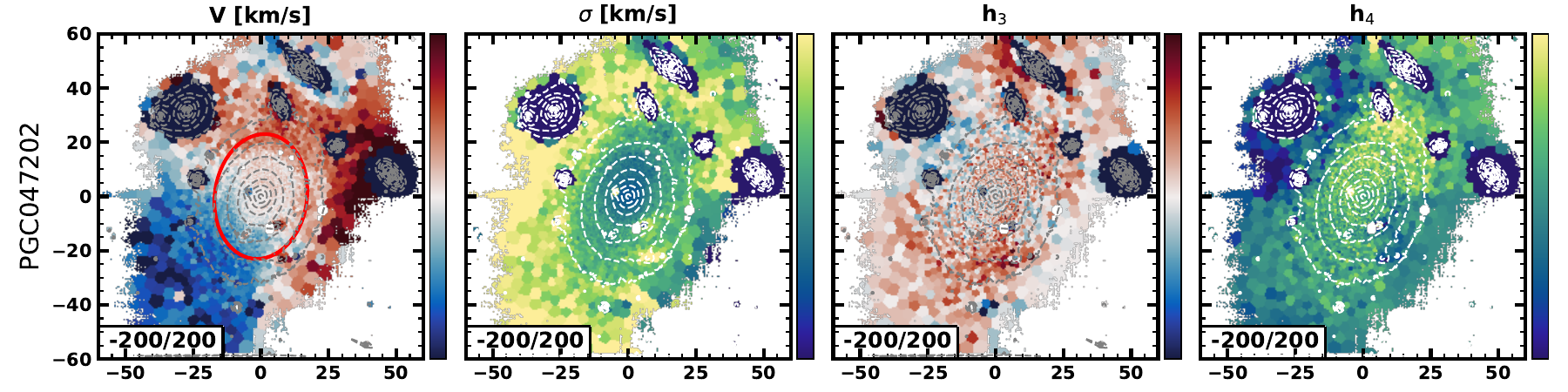} 
    \includegraphics[width=\textwidth]{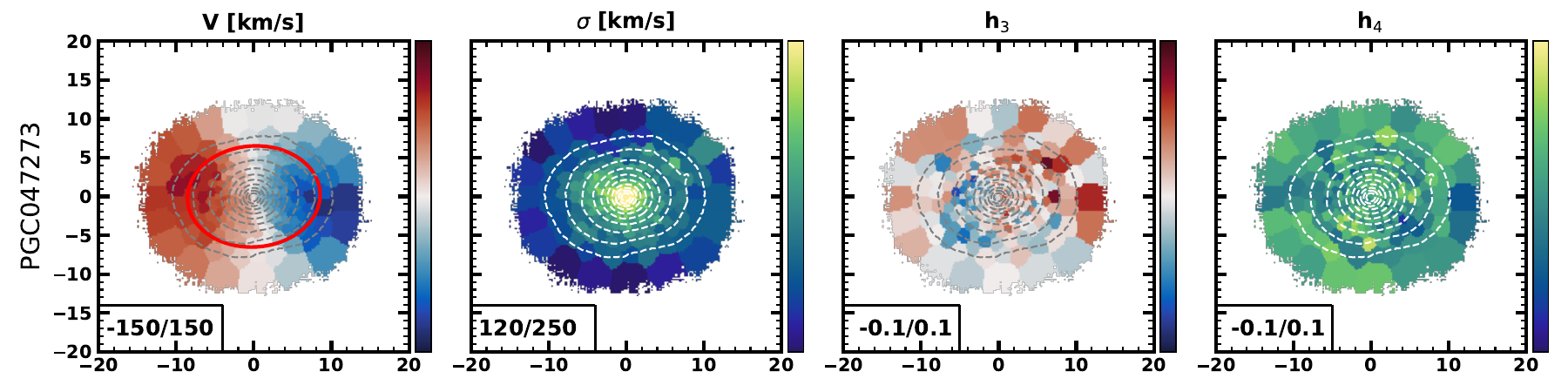} 
      \caption{Continued}
\end{figure*}

  \begin{figure*}
   \ContinuedFloat
    \includegraphics[width=\textwidth]{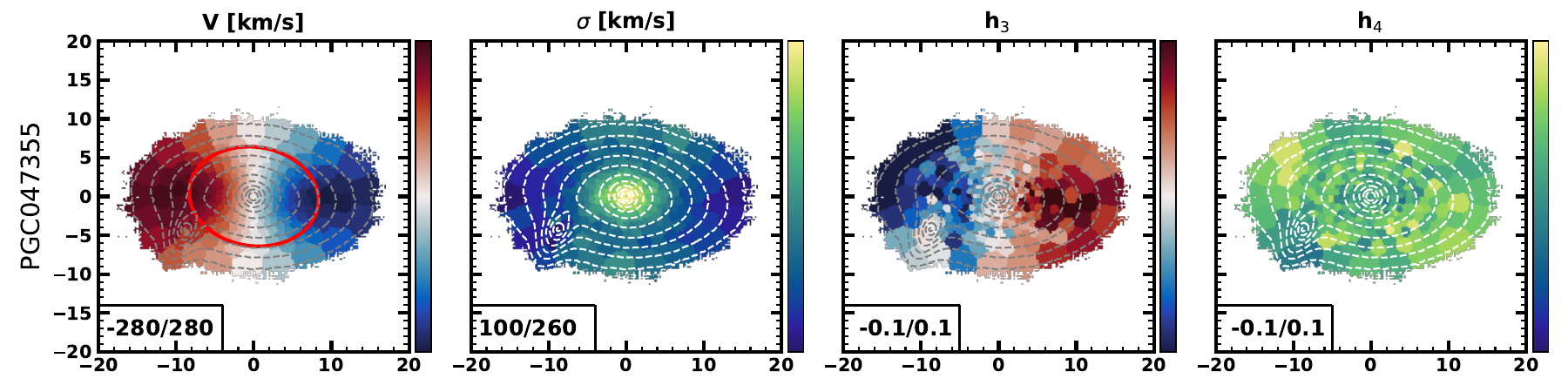} 
    \includegraphics[width=\textwidth]{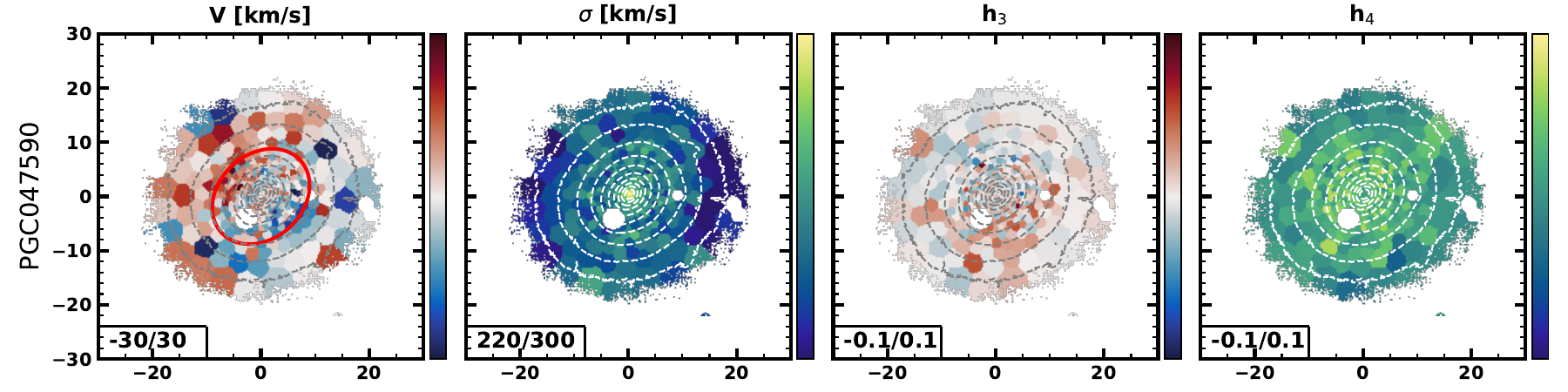} 
    \includegraphics[width=\textwidth]{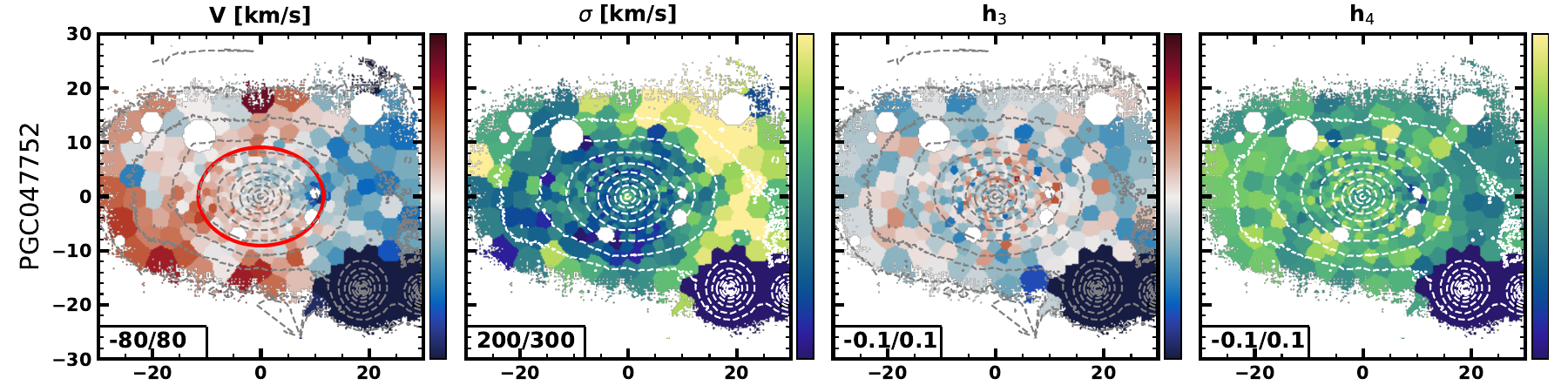} 
    \includegraphics[width=\textwidth]{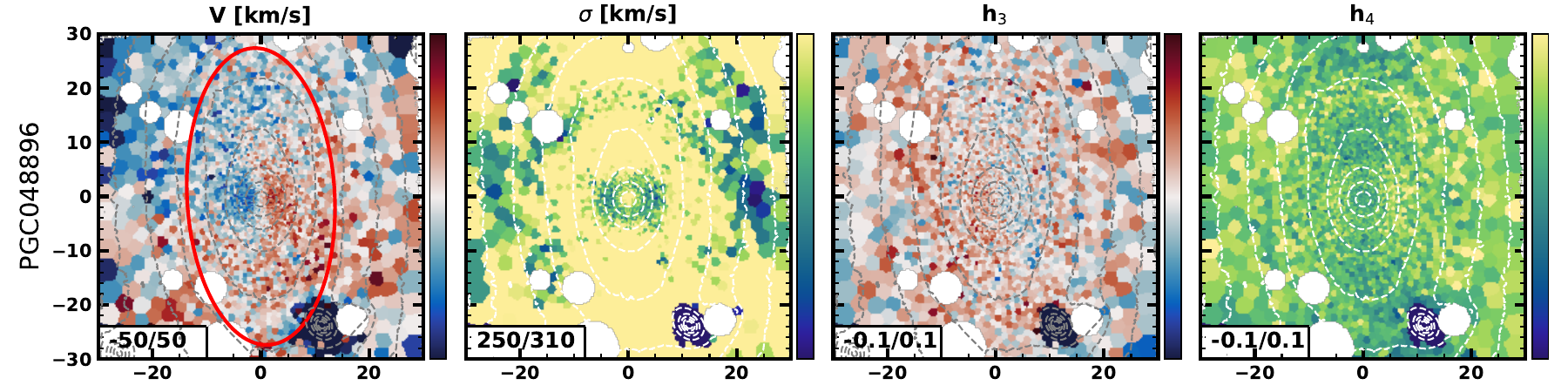} 
    \includegraphics[width=\textwidth]{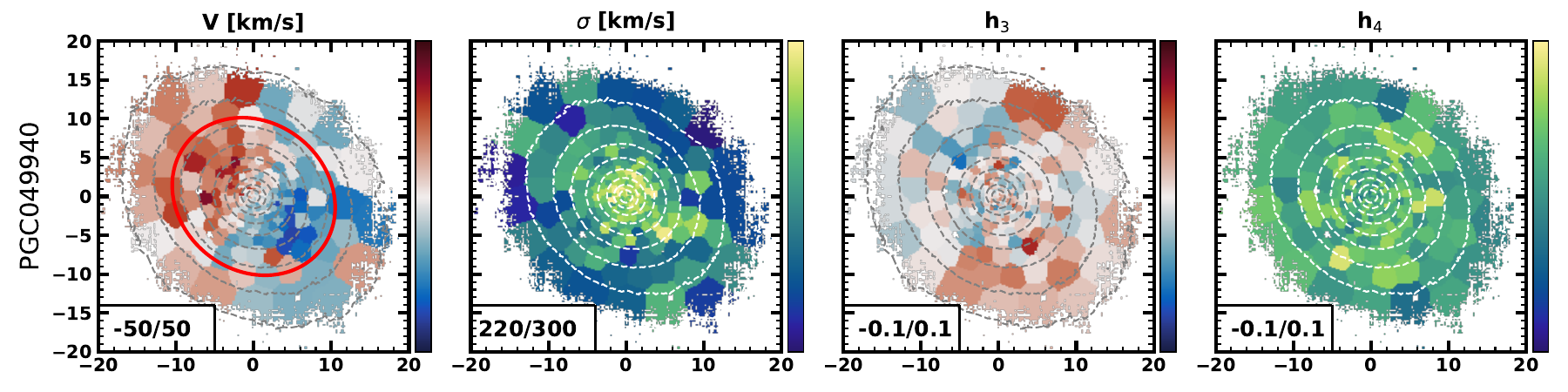} 
      \caption{Continued}
   \end{figure*}

  \begin{figure*}
   \ContinuedFloat
    \includegraphics[width=\textwidth]{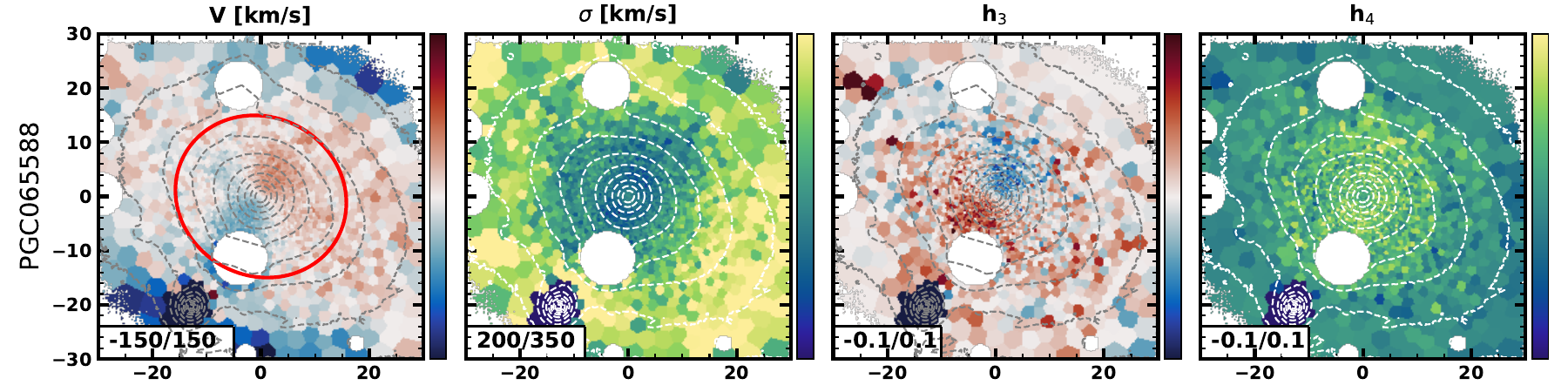} 
    \includegraphics[width=\textwidth]{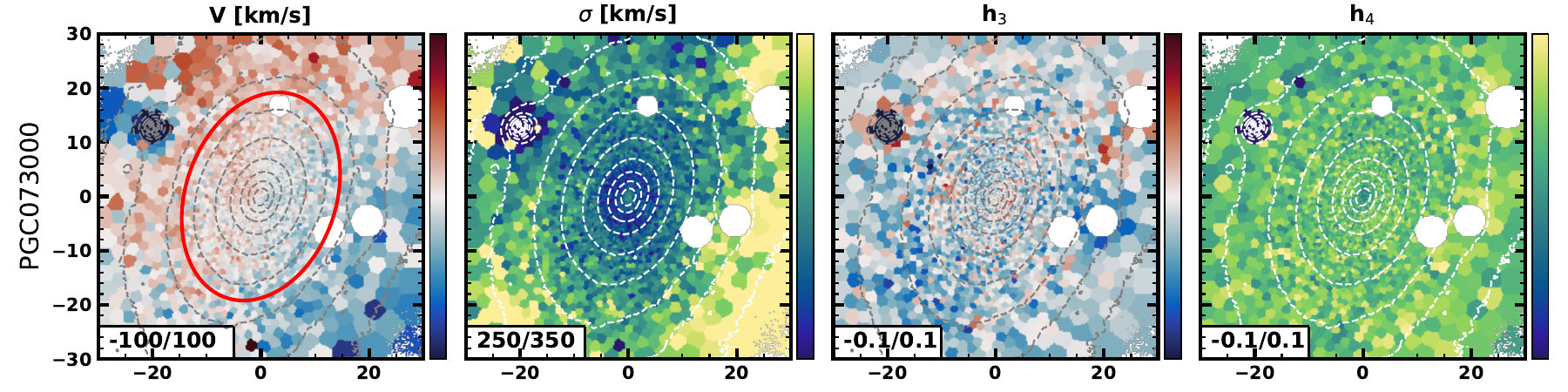} 
    \includegraphics[width=\textwidth]{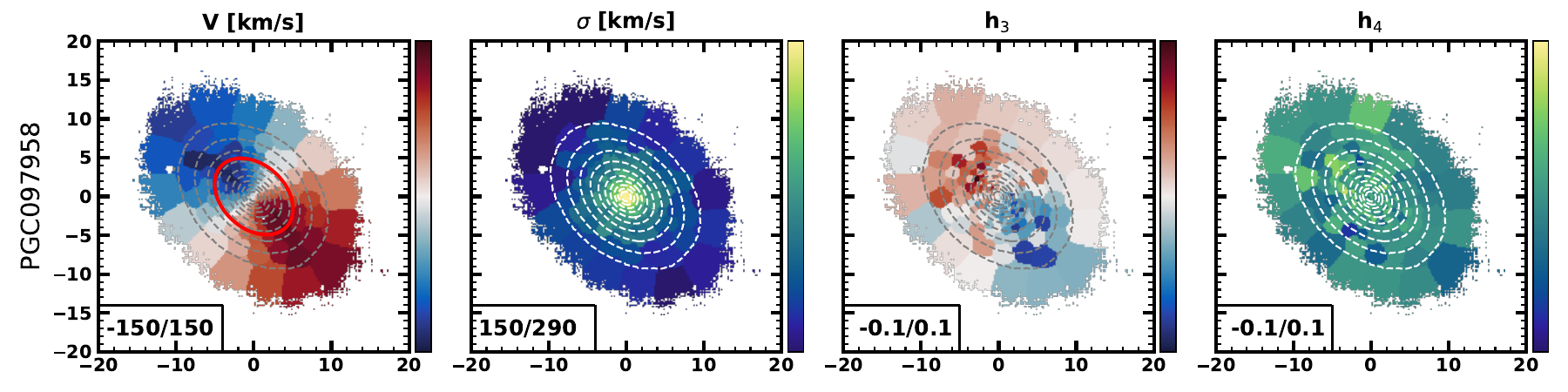} 
    \includegraphics[width=\textwidth]{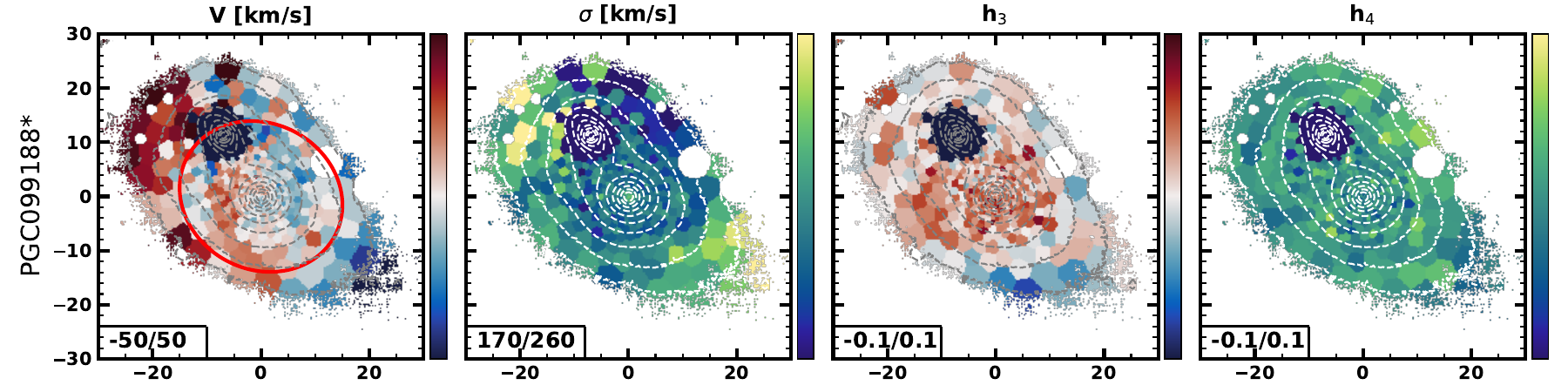} 
    \includegraphics[width=\textwidth]{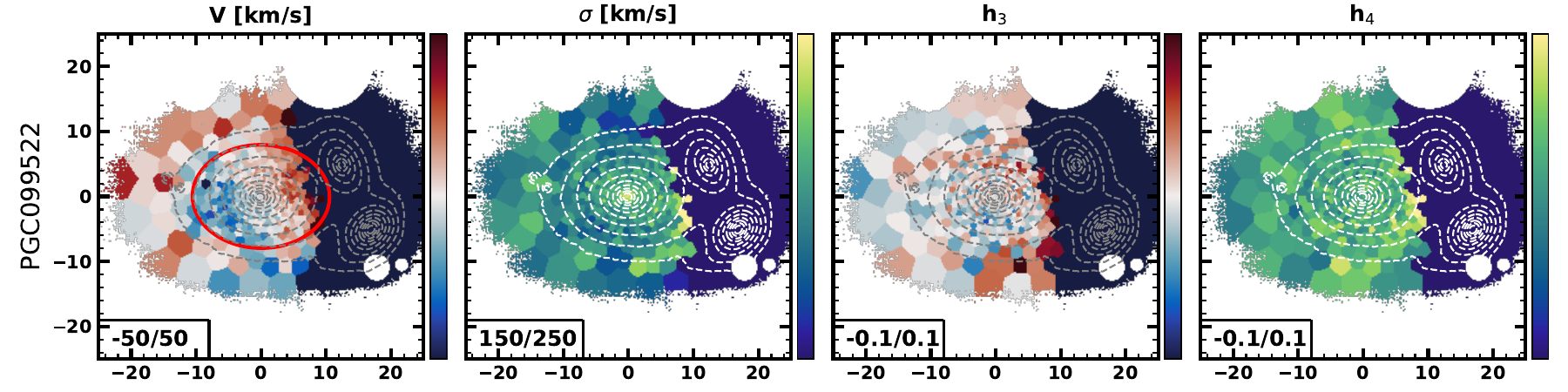} 
      \caption{Continued}
   \end{figure*}


\end{appendix}

\end{document}